\documentclass[a4paper,floatfix,rmp,twocolumn,showkeys,superscriptaddress]{revtex4}
  \usepackage[latin1]{inputenc}
  \usepackage{graphicx}
  \usepackage{rotating}
  \usepackage{mathtools}
  \setcitestyle{numbers,square,comma}
  \usepackage{diagbox}
  \usepackage{makecell}
  \newcolumntype{?}{!{\vrule width 1pt}}
  \newcolumntype{/}{!{\vrule width 2pt}}

  \usepackage{amsmath}
  \DeclareMathOperator*{\argmax}{arg\!max}

\begin{document}

  \title{Games in rigged economies}

  \providecommand{\CNB}{Departamento de Biolog\'ia de Sistemas, Centro Nacional de Biotecnolog\'ia (CSIC), C/ Darwin 3, 28049 Madrid, Spain. }
  \providecommand{\GISC}{Grupo Interdisciplinar de Sistemas Complejos (GISC), Madrid, Spain. }
  \providecommand{\IFISC}{Instituto de F\'isica Interdisciplinar y Sistemas Complejos IFISC (CSIC-UIB), Palma de Mallorca, Spain. }
  
  \author{Lu\'is F Seoane}
    \affiliation{\CNB}
    \affiliation{\IFISC}

  \vspace{0.4 cm}
  \begin{abstract}
    \vspace{0.2 cm}

    Modern economies evolved from simpler human exchanges into very convoluted systems. Today, a multitude of aspects
    can be regulated, tampered with, or left to chance; these are economic {\em degrees of freedom} which together shape
    the flow of wealth. Economic actors can exploit them, at a cost, and bend that flow in their favor. If intervention
    becomes widespread, microeconomic strategies of different actors can collide or resonate, building into
    macroeconomic effects. How viable is a `rigged' economy, and how is this viability affected by growing economic
    complexity and wealth? Here we capture essential elements of `rigged' economies with a toy model. Nash equilibria of
    payoff matrices in simple cases show how increased intervention turns economic degrees of freedom from minority into
    majority games through a dynamical phase. These stages are reproduced by agent-based simulations of our model, which
    allow us to explore scenarios out of reach for payoff matrices. Increasing economic complexity is then revealed as a
    mechanism that spontaneously defuses cartels or consensus situations. But excessive complexity enters abruptly into
    a regime of large fluctuations that threaten the system's viability. This regime results from non-competitive
    efforts to intervene the economy coupled across degrees of freedom, becoming unpredictable. Thus non-competitive
    actions can result in negative spillover due to sheer economic complexity. Simulations suggest that wealth must grow
    faster than linearly with economic complexity to avoid this regime and keep economies viable in the long run. Our
    work provides testable conclusions and phenomenological charts to guide policing of `rigged' economic systems. 

  \end{abstract}


\maketitle

 \section{Introduction}
    \label{sec:1}

    The existence of `rigged' economic scenarios is amply acknowledged. Most notable examples are non-competitive
    markets \cite{Tirole1988, WilligSchmalensee1989}, legal or illegal, such as cartels, or natural monopolies
    \cite{LaffontTirole1993}. In these, all actors usually cooperate to secure similar profits. This entails
    `handcrafting' some aspects of the economic games in which they engage. In competitive markets we also find illegal
    schemes (e.g. inside trading) or innovative, often borderline legal, enterprises to explore unprecedented economic
    possibilities -- e.g. anticipating a broker's moves with faster internet cables \cite{Lewis2015}. Such
    out-of-the-box thinking is part of the economy's open-ended nature \cite{Schumpeter1942, Ferguson2008}. It redesigns
    the rules of the game and easily results in a sentiment that ``the market is rigged'' \cite{Lewis2015}. Even if all
    actors stick to the norms and do not innovate, competitive markets are strongly regulated. Some conditions (e.g.
    demanding a minimum equity to participate) are designed by governments or international institutions. They might
    change due to democratic consensus or lobbying. If powerful firms bend the rules systematically, regulatory capture
    happens \cite{Stigler1971, Tirole1986, LaffontTirole1991} threating democracy at large \cite{AcemogluRobinson2005,
    AcemogluRobinson2019}. As transnational markets grow ever more complex and faster, slow public bureaucracies might
    lag behind and abdicate into nimbler private regulators \cite{Draghi2019, Barucca2020}. 

    Through and through, economies are `rigged'. Available games are somehow manufactured. Once established, they remain
    open to manipulations that might i) impact costs and rewards of economic games, ii) cap the information available,
    or iii) limit the number of players allowed to partake. This can be achieved through publicity, bribes, threats,
    imposing tariffs, etc. More abstractly, we can think of {\em degrees of freedom} that can be harnessed in economic
    systems. Each degree of freedom is a pocket of opportunity that can be exploited, contested or uncontested, at some
    cost. Envelop theorems assess changes of likely payoffs when a game is altered externally \cite{MilgromShannon1994,
    Caputo1996, AcemogluJensen2013}. Wolpert and Grana \cite{WolpertGrana2019} recently wondered how much an agent
    should pay if she (and no other actor involved) was given this control before playing a game. The decision boils
    down to a positive payoff balance {\em with} versus {\em without} intervention. Thus a single agent is offered
    control, at a cost, over a single economic degree of freedom.
    
    Here we study what happens when multiple actors are allowed, also at a cost, to manipulate several economic degrees
    of freedom. Different efforts might align or not, yielding uncertain returns. A single agent's decision to rig one
    game (as per \cite{WolpertGrana2019}) might be of limited consequence in isolation. But effects may be amplified,
    mitigated, or produce emergent phenomena when coupled across games and players. We are interested in how microscopic
    fates scale up to macroeconomic trends, so we adopt a systemic perspective. More available degrees of freedom result
    in more complex economies -- intervention possibilities grow combinatorially and more external variables become
    relevant if extra degrees of freedom are left unchecked. How do system-wide dynamics of a rigged economy depend on
    its complexity? How much can such economies grow? Open-endedly, perhaps? Do they collapse, unable to sustain their
    participants? How is this affected by the amount of wealth generated and distributed? What is a natural level of
    intervention depending on these aspects? 

    \begin{figure} 
      \begin{center} 
        \includegraphics[width=\columnwidth]{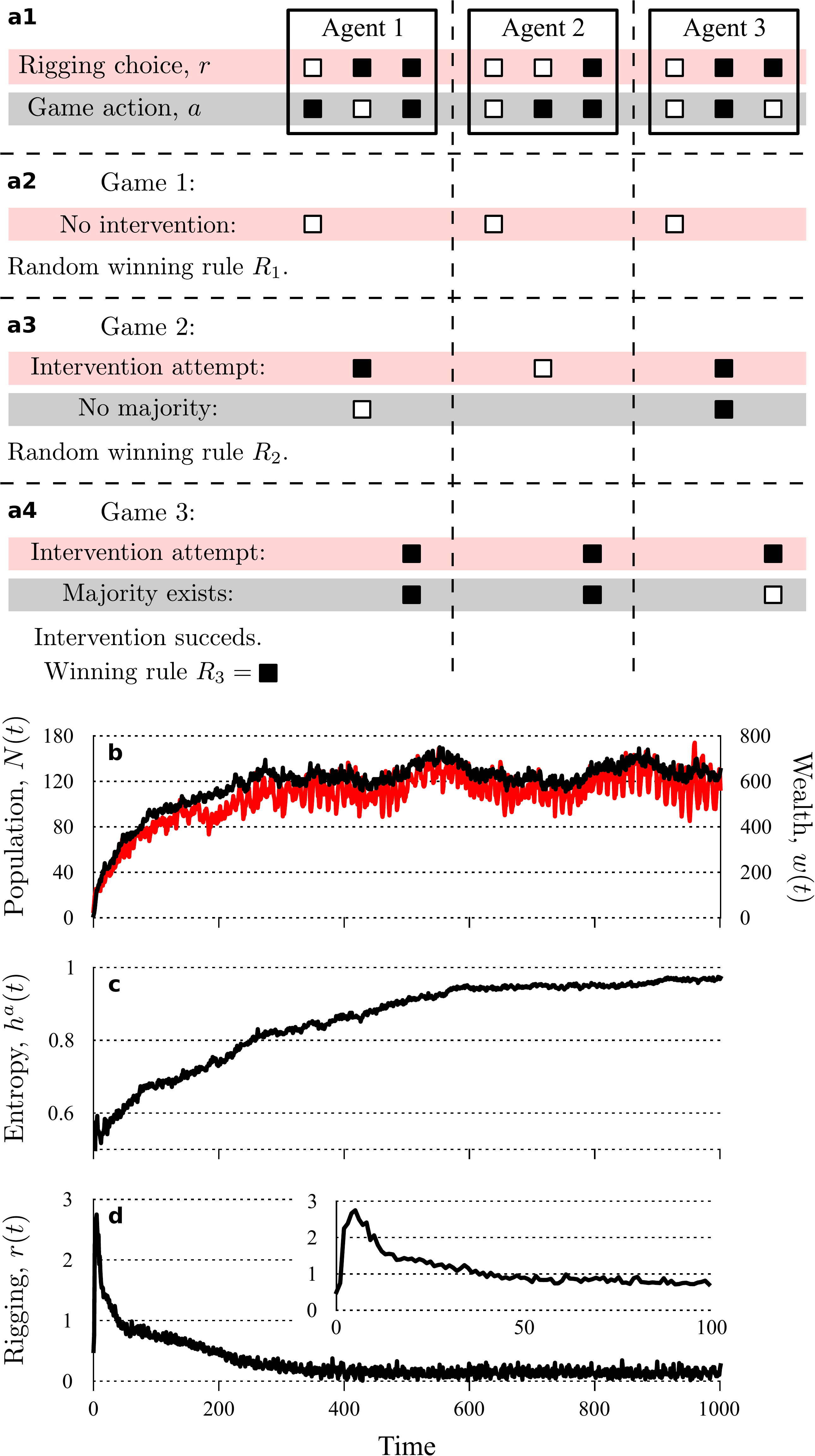}

        \caption{{\bf A rigged economy and its dynamics.} {\bf a} Three agents choose (red shade) whether to rig each
game (black boxes) or not (empty boxes), and (gray shade) what to play in each game ($a^i_k=1$ or $0$). {\bf b-d}
Dynamics of model measurements for $C_C = 10$, $n = 20$, and $B = 100$ (thus $b = 5$). {\bf b} Population (black, left
scale) and wealth (red, right scale). }

        \label{fig:1}
      \end{center}
    \end{figure}

    We could tackle these questions rigorously through utility functions that discount intervention costs, as in
    \cite{WolpertGrana2019}, extended to multiple agents and games; but this quickly becomes untreatable. Instead,
    inspired by agent-based models and complex adaptive systems \cite{Gintis2000, Whitehead2008, Buchanan2008,
    Sornette2017, DevittLeeBoghosian2018, LiLi2019}, we capture essential elements that affect our research questions
    into a toy economy. We assume a population of agents who engage in $n$ economic games. Each game has a rule that
    randomly determines its winning strategy. Agents can pay to intervene each game's rule, affecting the winning
    strategy for all (figure \ref{fig:1}{\bf a}). These games constitute our {\em degrees of freedom}, thus $n$ is a
    proxy for the economy's complexity. An amount of wealth, $B$, is distributed among winners. $B$ reminds us of a GDP
    and is a proxy for our economy's size. The model is described in detail in section \ref{sec:2}. We miss some
    important ways of manipulating real economies -- such shortages and model extensions are discussed in section
    \ref{sec:4}. 

    We write complete payoff matrices for some simple scenarios. Their analysis (section \ref{sec:3.1}) shows that
    increasing intervention switches isolated degrees of freedom from minority to coordination games. In between, Nash
    equilibria are mixed strategies, anticipating dynamic struggles. We explore increasing economy size and complexity
    with simulations based on agents of bounded rationality and Darwinian dynamics to select successful strategies. We
    argue (section \ref{sec:4}) that our results should not depend critically on the agent's rationality and the
    Darwinism. We simulate model dynamics for a small, fixed number of degrees of freedom as economy size grows (section
    \ref{sec:3.2}). This reveals the same progression: from minority, through dynamic, to coordinating regimes. The
    later remind us of cartels. Adding degrees of freedom abruptly halts within-game coordination, suggesting an
    empirical test: increased economic complexity should dissolve cartels spontaneously. We study our toy economy's
    viability as its complexity grows large and its size scales appropriately (section \ref{sec:3.3}). Economies whose
    size does not grow fast enough with their complexity fall in a large-fluctuations regime that threatens their
    viability -- thus non-competitive actions can have negative spillovers as agents and degrees of freedom become
    coupled en masse. Our toy model allows us to find limit regimes (e.g. within-game coordination, large-fluctuations,
    etc.) that emerge from essential elements potentially common to any `rigged' economy. We lay out comprehensive maps
    of such regimes (section \ref{sec:3.4}). Their occurrence is tied to a few, yet abstract parameters. In section
    \ref{sec:4} we speculate how we might link them to real-world economies.

  \section{Methods} 
    \label{sec:2}

    \subsection{Model description}
      \label{sec:2.1}
  
      Our toy economy (Figure \ref{fig:1}{\bf a}) consists of a fixed number of games, $n$; and a population of $N(t)
      \in [0, N^{max}]$ agents that changes over time. At each iteration, every agent has to play all games, which admit
      strategies $0$ or $1$. The strategies played by agent $A^i$ are collected in an array: $a^i \equiv [a^i_k, k=1,
      \dots, n]$. Besides, a second array $r^i \equiv [r^i_k, k=1, \dots, n]$ codifies whether $A^i$ attempts to rig
      game $k$ ($r^i_k = 1$) or not ($r^i_k = 0$). The combination $(a^i_k, r^i_k)$ constitutes the proper strategy of
      agent $A^i$ towards game $k$. However, to aid the model's discussion, we use the word `strategy' only to name
      $a^i_k$. 
  
      At each iteration a rule exists, common to all agents, that determines the winning strategy for each game: $R(t)
      \equiv [R_k(t)\in\{0, 1\}, k=1, \dots, n]$. If any agents attempt to rig game $k$, $R_k(t)$ takes the most common
      action among those rigging agents (Figure \ref{fig:1}{\bf a4}): 
        \begin{eqnarray}
          R_k(t) &=& \argmax_{\bar{a}\in\{0,1\}} \left(||\left\{ A^i, a^i_k = \bar{a}, r^i_k=1 \right\}|| \right). 
        \end{eqnarray}
      In case of draw (including no intervention), $R_k(t)$ is set randomly (Figure \ref{fig:1}{\bf a2-3}). Each agent
      pays an amount $C_R$ for each intervention attempt -- successful or not. If $A^i$ has a wealth $w^i(t)$ at the
      beginning of a round, after setting $R(t)$ this becomes:
        \begin{eqnarray}
          w^i(t + \Delta t_{rig}) &=& w^i(t) - C_R\sum_k r^i_k. 
        \end{eqnarray}
  
      Each round, an amount $b$ is ruffled at each game -- a total wealth $B = nb$ is potentially distributed. The
      amount allocated to game $k$ is split between all agents who played the winning strategy, $R_k(t)$. After this: 
        \begin{eqnarray}
          w^i(t + \Delta t_{play}) &=& w^i(t + \Delta t_{rig}) \nonumber \\ 
          && + \> b\sum_{k} {\delta(a^i_k, R_k(t)) \over N^w_k(t)}, 
        \end{eqnarray}
      where $\delta(\cdot, \cdot)$ is Kronecker's delta and $N^w_k(t)$ is the number of winners of game $k$. If $w^i(t +
      \Delta t_{play}) < 0$, the $i$-th agent is removed, decreasing the population by $1$. 
  
      If $w^i(t + \Delta t_{play}) > C_C$, $A^i$ has a child and an amount $C_C$ is subtracted from $w^i$. A new agent
      is generated which inherits $a^i$ and $r^i$. Each of the bits in these arrays flips once with a probability
      $p_\mu$. After this, both arrays remain fixed throughout the new agent's lifetime. We generate an integer number
      $j \in [1, N^{max}]$ to allocate the new individual. If $j \le N(t)$, the new agent becomes $A^j$. The former
      agent in that position is removed, its wealth is lost, and the population size remains unchanged. If $j > N(t)$,
      the new individual is appended at the end of the pool and the population grows by $1$.

    \subsection{Measurements on model dynamics}
      \label{sec:2.2}

      For each simulation we set model parameters ($C_R = 1$, $C_C=10$, $p_\mu = 0.1$, and $N^{max} = 1000$; but
      variations are explored in Appendix \ref{app:3} to show the generality of our results). We explore ranges of $n$
      and $B$ to address the main questions -- i.e. ``how do rigged economies behave as their complexity and size
      change?'' 

      Model simulations start with a single agent $A^1$ with random strategies and no interventions ($r^1_k = 0, \forall
      k$). In average, $A^1$ accrues half of the distributed wealth until $w^1(t) > C_C$. As new descents fill the
      population, reinforcing or competing strategies unfold. After a rapid initial growth, population size and wealth
      reach an attractor (Figure \ref{fig:1}{\bf b}). We asses these attractors numerically. Simulations run for $5000$
      iterations. We take averages (noted $\left<\cdot\right>$) of diverse quantities over the last $500$ iterations.
      For example, population size $\left<N\right>$, for which we also report normalized fluctuations
      $\sigma(N)/\left<N\right>$, where $\sigma(\cdot)$ indicates standard deviation. Unless $B$ is very small, $5000$
      iterations suffice to observe convergence (Supporting Figures \ref{fig:SI2}, \ref{fig:SI4}, and \ref{fig:SI6}).
      
      To measure the heterogeneity of strategies in the population, we take the fraction $f_k(t)$ of agents with
      $a_k=1$:
        \begin{eqnarray}
          f_k(t) = \sum_{i=1}^{N(t)} {\delta(a^i_k, 1) \over N(t)}, 
        \end{eqnarray}
      from which we compute the entropy: 
        \begin{eqnarray}
          h^a_k(t) &=& - \Big[f_k(t)log_2(f_k(t)) \nonumber \\ 
          && + (1-f_k(t))log_2(1-f_k(t)) \Big] 
        \end{eqnarray}
      and mean entropy across games (Figure \ref{fig:1}{\bf c}): 
        \begin{eqnarray}
          h^a(t) &=& {1 \over n} \sum_{k=1}^n h^a_k(t). 
        \end{eqnarray}
      If $h^a_k(t) = 0$, all agents play the same strategy in game $k$. This quantity is maximal ($h^a_k(t)=1$) if the
      population splits in half around that game. If $h^a(t) = 0$, agents play the same strategy in each game, but not
      necessarily the same one across games. If $h^a(t)=1$, agents are split in half at each game, but this split is not
      necessarily consistent across games. 

      Finally, we introduce the {\em rigging pressure} on a game: 
        \begin{eqnarray}
          r_k(t) = {1 \over N(t)} \sum_{i=1}^{N(t)} r^i_k, \>\>\>\> r_k(t) \in [0, 1]; 
        \end{eqnarray}
      as well as the total rigging pressure (Figure \ref{fig:1}{\bf d}): 
        \begin{eqnarray}
          r(t) = \sum_{k=1}^n r_k(t), \>\>\>\> r(t) \in [0, n]; 
        \end{eqnarray}
      and average rigging pressure per game $r(t)/n \in [0, 1]$.

  \section{Results} 
    \label{sec:3}

    \subsection{Intervention turns minority into majority games}
      \label{sec:3.1}

      \begin{table*}
        \includegraphics[width=0.8\textwidth]{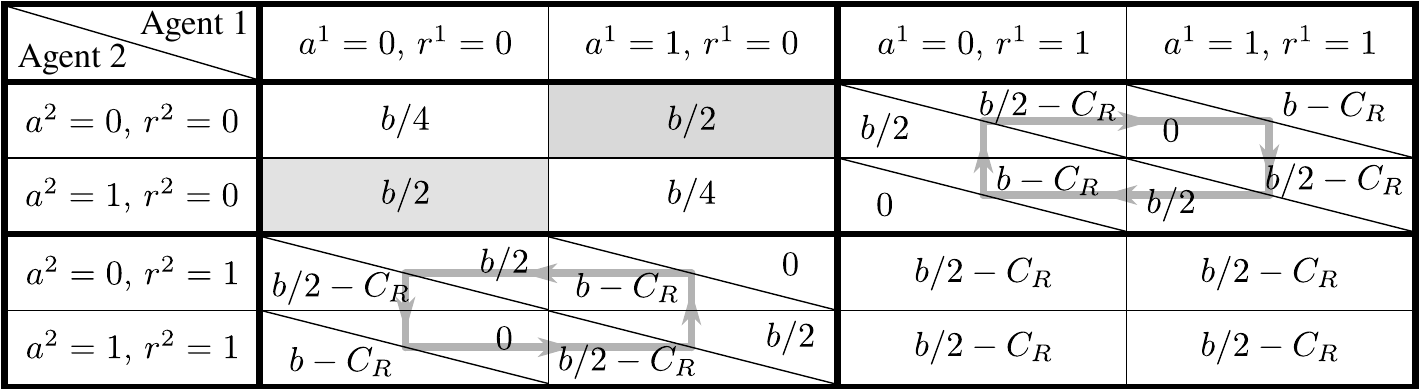}

        \caption{{\bf Payoff matrix of one game with two players. } Table entries are labeled by each agent's strategy
$a^1=0,1$ and rigging choice $r^1=0,1$. Each cell displays average payoff with no death or reproduction for fixed
options. Gray cells are Nash equilibria if $C_R > b/2$. Gray circuits indicate possible dynamic situations that emerge
for $C_R < b/2$. } 

        \label{tab:1}
      \end{table*}

      Before looking at model dynamics we can gain some insight from payoff matrices in simple cases. Population size
      affects these matrices: earnings are split among winners; and more agents imply more distinct, possible
      correlations between strategies and rigging choices. Hence, utility functions rapidly become very complex. In
      appendix \ref{app:1} we discuss payoff matrices for a single game and one player (Supporting Table \ref{tab:ST1})
      and for one game and three players (Supporting Tables \ref{tab:ST3}, \ref{tab:ST4}, and \ref{tab:ST5}). All
      matrices show average earnings over time if strategies, rigging choices, and population size are fixed. 

      Table \ref{tab:1} presents the payoff matrix for one game with two players. If $C_R > b/2$, rigging the game is
      prohibitive. Then, the system has the Nash equilibria marked in gray -- both agents try to take opposite actions
      $a^1 \ne a^2$. With no intervention, we deal with a minority game. For larger populations it pays even more to be
      in the minority (Supporting Table \ref{tab:ST3}). These equilibria disappear if intervention is cheap enough ($C_R
      < b/2$ in table \ref{tab:1}, but depends on population size). Then, it becomes more profitable for one player to
      rig the game while playing a minority strategy. But it also becomes better, for the other agent, to parasitize the
      other's effort -- turning the winning strategy into a majority. If the game were played sequentially, a dynamic
      scenario ensues alternating minority and majority configurations (gray circuits in table \ref{tab:1}). As more
      players are added, the stakes become higher and the situation more complex. Each player wants to be in the
      majority among rigging agents, but in the minority among non-rigging ones. For $n>3$, if all agents are
      intervening (red frame, Supporting Table \ref{tab:ST5}), the sub-game's Nash equilibrium is a full coordination.
      This is not a global equilibrium, but large coordinations emerge in our simulations for rising intervention levels
      (see next section). Note how all agents rigging a game in an agreed-upon way to share profits resembles a cartel. 

      Payoff matrices are equal for all games. If $n$ games were played in isolation (i.e. wealth earned by manipulating
      a game could not be invested into another), we would observe the same transition to within-game coordination for
      each degree of freedom as intervention takes hold. What happens when we lift such compartmentalization?

    \subsection{Fixed complexity and growing wealth}
      \label{sec:3.2}

      We now study model dynamics and stability for a fixed number of games and varying economy size. Discussion of the
      rich behavior uncovered follows in Appendix \ref{app:2}. 

      \begin{figure*}
        \begin{center}
          \includegraphics[width=\textwidth]{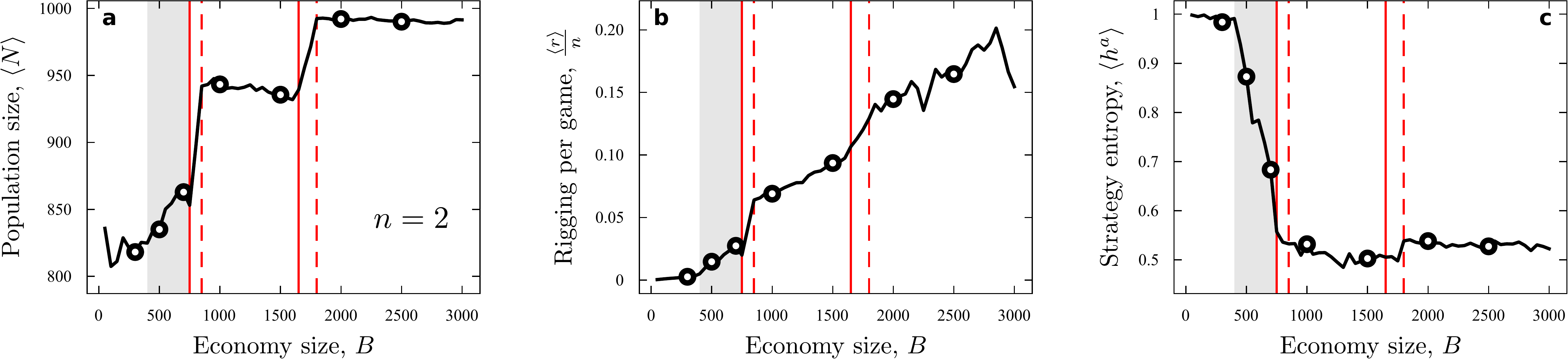}
  
        \caption{{\bf Model behavior for $n=2$ games and growing wealth, $B$.} {\bf a} $\left< N \right>$, {\bf b}
$\left<r\right> / n$, and {\bf c} $\left<h^a\right>$. Open circles indicate $B$ values for which we plot sample dynamics
in Supporting Figure \ref{fig:SI4}. Red vertical lines loosely indicate discrete jumps in $\left< N \right>$. }
  
          \label{fig:4}
        \end{center}
      \end{figure*}

      Figure \ref{fig:4}{\bf a} shows $\left<N\right>$ for $n=2$ games. Circles over the plots indicate values of $B$
      for which a stretch of the dynamics is plotted in Supporting Figure \ref{fig:SI4}. Generally, $\left<N\right>$
      increases with the economy size -- i.e. as more money becomes available to sustain more agents or to invest into
      rigging more games. Indeed, the rigging pressure per game (Figure \ref{fig:4}{\bf b}) grows more or less
      monotonously. $\left<N\right>$ is not so parsimonious. For roughly $B < 750$ it grows steadily. At $B \sim 750$ it
      jumps swiftly, then remains similar but slightly declining up until $B \sim 1600$, when it undergoes another
      abrupt increase. 
      
      These population boosts seem associated to varying coordination. Figure \ref{fig:4}{\bf c} shows that the strategy
      entropy $\left<h^a\right>$ drops sharply before the first boost (shaded area). Before that drop, resources are
      scarce and rigging the economy is difficult. Either strategy is equally likely to win, so agents playing either
      option are equally abundant (Supporting Figure \ref{fig:SI4}{\bf a}). As $B$ grows, more resources become
      available to rig the games. Either $1$ or $0$ becomes the winning strategy over longer time stretches, resulting
      in temporary selective preferences for one strategy over the other, and oscillatory dynamics ensue (Supporting
      Figure \ref{fig:SI4}{\bf b-c}). As $\left<h^a\right>$ falls definitely, agents coordinate their strategies
      (Supporting Figure \ref{fig:SI4}{\bf d-e}). These dynamics shifts happen simultaneously in all games -- as if, so
      far, payoff matrices were essentially independent for each degree of freedom. By $B \sim 1000$ we exhausted all
      within-game regimes uncovered in payoff matrices: from minority games, through a dynamic struggle, to a mostly
      majority game. The final population boost at $B \sim 1600$ must entail emerging correlations across games -- e.g.
      clustering agents that play the minority vs majority strategies in both games. 

      Supporting Figures \ref{fig:SI6.5} and \ref{fig:SI7} compare $\left<N\right>$, $\left<r\right>/n$, and
      $\left<h^a\right>$ as economy size grows for different, fixed $n$. With more games, more discrete jumps in
      $\left<N\right>$ appear. These arise, potentially, from the combinatorially growing coordination possibilities
      across games. They happen after the oscillatory phases (Supporting Figures \ref{fig:SI5} and \ref{fig:SI6}{\bf
      c-e} for $n=3$). This again suggests that within-game coordination happens first, simultaneously for all games;
      then degrees of freedom start coupling with each other. Some regimes have similar $\left<N\right>$ for different
      $n$ (horizontal dashed lines, Supporting Figure \ref{fig:SI6.5}{\bf a}), suggesting that they are effectively
      similar. Population boosts succeed each other more rapidly for larger $n$, approaching a continuous buildup
      instead of discrete jumps (Supporting Figure \ref{fig:SI7}{\bf a}). This is not reflected by $\left< h^a \right>$,
      which only drops once due to within-game coordination. The $\left<h^a\right>$ plateau is higher for larger $n$
      (Supporting Figure \ref{fig:SI7}{\bf c}), indicating that across-game correlations weaken or interrupt within-game
      coordination. Above, we compared such coordination to cartels: games are consensually rigged to favor most actors.
      Our results suggest that increasing economic complexity prevents the formation of such consensus, defusing
      cartels, even with rising intervention levels. This is a testable conclusion of our model.

    \subsection{Growing wealth and economic complexity}
      \label{sec:3.3}

      \begin{figure*}[t]
        \begin{center} 
          \includegraphics[width=0.9\textwidth]{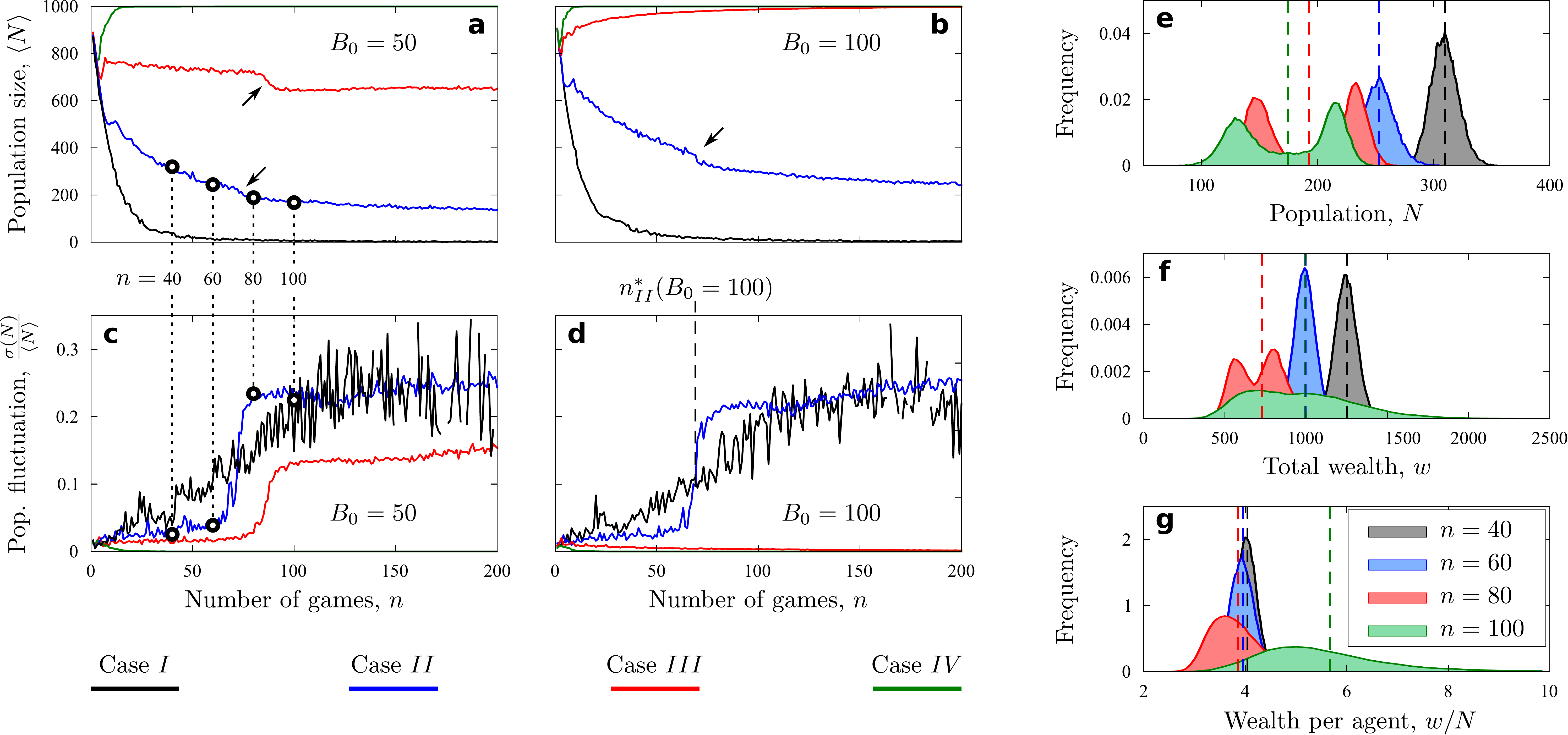}
  
          \caption{{\bf Population size and fluctuations for growing $n$ and $B \equiv B(B_0, n)$.} {\bf a-b} show
$\left< N \right>$ and {\bf c-d} show $\sigma(N) / \left<N\right>$. $B_0 = 50$ in {\bf a}, {\bf c}; $B_0 = 100$ in {\bf
b}, {\bf d}. Arrows mark the onset of large-fluctuations in $\left< N \right>$. Dotted, vertical lines and circles in
{\bf a}, {\bf c} indicate $n$ values examined in {\bf e-g}. {\bf e} Probability density functions of population size, $N$;
{\bf f} total wealth, $w$; and {\bf g} wealth per capita $w/n$. }
  
          \label{fig:2}
        \end{center}
      \end{figure*}

      We now change the number of games as wealth scales as $B = B(B_0, n)$. The constant $B_0$ is a normalizing factor
      to facilitate comparisons. We explore four cases: 
        \begin{itemize}
  
          \item[$I$:] A fixed wealth $B_{I} = B_0$ is split evenly between all games: $b_{I} = B_0/n$. Returns per game
drop as the economy becomes more complex. 

          \item[$II$:] Each game distributes a fixed amount $b_{II} = B_0$, total wealth grows linearly $B_{II} = B_0
\cdot n$. Returns per game remain constant against growing complexity. 
  
          \item[$III$:] Each degree of freedom {\em revalues previously existing games} logarithmically: $b_{III} = B_0
\left(\log(n) + 1\right)$. Total wealth grows as $B_{III} \sim B_0 \cdot n \left(\log(n) + 1\right)$. 

          \item[$IV$:] Each degree of freedom {\em revalues previously existing games} linearly: $b_{IV} = B_0 \cdot n$.
Total wealth grows quadratically: $B_{IV} = B_0 \cdot n^2$. 
  
        \end{itemize}

      Figure \ref{fig:2}{\bf a-b} shows $\left< N \right>$ for each scenario. Extreme cases $I$ (black curves) and $IV$
      (green) are relatively uninteresting: Stable population size declines quickly for $I$. As the economic complexity
      grows and returns per game drop, more intervention is needed to secure the same earnings. Such rigged economies
      collapse if they become too complex. For $IV$, wealth grows so quickly with $n$ that, promptly, population
      saturates.

      Intermediate cases $II$ (blue curves) and $III$ (red) are more interesting. With $B_0 = 50$ (Figure
      \ref{fig:2}{\bf a}) and $B_0=100$ (Figure \ref{fig:2}{\bf b}), $\left< N \right>$ in declines slowly for $II$.
      Thus, in general, a rigged economy's wealth must grow faster than linearly with its complexity to remain viable.
      In case $III$, $\left< N \right>$ saturates for $B_0 = 100$; but not for $B_0 = 50$, for which population seems
      stagnant. 

      For case $II$, and $III$ with $B_0 = 50$, fluctuations in population size reveal the existence of thresholds,
      $n^*_{II/III}(B_0)$, at which system dynamics change abruptly (Figure \ref{fig:2}{\bf c-d}). This affects
      $\left<N\right>$ marginally (arrows in Figure \ref{fig:2}{\bf a-b}), but the increase in $\sigma(N) / \left< N
      \right>$ is always salient. For $n < n^*$, fluctuations are small ($< 5\%$). For $n > n^*(B_0)$ large fluctuations
      ($\sim 25\%$ for case $II$ and $\sim 15\%$ for case $III$) set in. There is an absorbing state at $N(t)=0$, thus
      fluctuations of $15 - 25\%$ system size can compromise its viability. 

      We explore the transition to large fluctuations by simulating case $II$ below ($n=40, 60$) and above ($n=80, 100$)
      their onset at $n^*$. We ran the model for $5000$ iterations and discarded the first $1000$. Figure
      \ref{fig:2}{\bf e} shows the probability of finding the system with population $N$. Below $n^*$ we see a neat
      Gau\ss ian; above, the distribution presents two balanced modes. Transitions between them contribute to the large
      fluctuations. We also plot total wealth (Figure \ref{fig:2}{\bf f}) and wealth per agent (Figure \ref{fig:2}{\bf
      g}). Their averages fall, first, and grow, eventually, as $n$ increases. Below $n^*$ clear Gau\ss ians appear
      again. Above, we observe broad tails, indicating inequality. Despite risking collapse, the average agent can be
      wealthier in the large-fluctuations regime.

    \subsection{Charting rigged economies} 
      \label{sec:3.4} 
      
      \begin{figure*}
        \begin{center}
          \includegraphics[width=\textwidth]{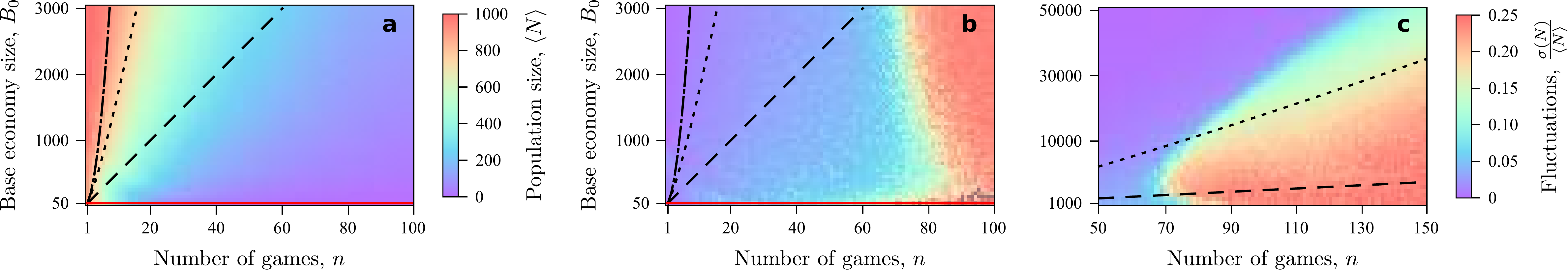}
  
          \caption{{\bf Comprehensive maps of $\left< N \right>$ and $\sigma(N) / \left< N \right>$. } Maps result from
simulating case $I$ (i.e. $B_I = B_0$, $b_I = B_0/n$) for ranges of economic complexity, $n$, and distributed wealth,
$B_0$. Black curves represent trajectories $B = B(B_0=50, n)$ for cases $II$, $III$, and $IV$ (dashed, dotted, and
dot-dashed respectively). Horizontal red lines at the bottom of maps {\bf a-b} represent case $I$ with $B_0 = 50$. {\bf
a} shows $\left< N \right>$ and {\bf b-c} show $\sigma(N) / \left< N \right>$ over two ranges of $n$ and $B_0$. }
  
          \label{fig:5}
        \end{center}
      \end{figure*}

      We run simulations of case $I$ ($B_{I} \equiv B_0$, $b_{I} \equiv B_0/n$) for ranges of economic complexity, $n$,
      and distributed wealth, $B_0$. This renders maps (Figure \ref{fig:5}) where the four scalings above can be read as
      curved sections. Trivially, case $I$ traces a horizontal line (solid, red; bottom of each map). Case $II$ traces a
      line with slope $B_0$ (dashed black lines). Cases $III$ (dotted black) and $IV$ (dash-dotted black) trace curves
      growing faster than linearly. Reading $\left<N\right>$ (Figure \ref{fig:5}{\bf a}) or $\sigma(N) / \left< N
      \right>$ (Figure \ref{fig:5}{\bf b-c}) along such curves renders the plots from figures \ref{fig:2}{\bf a-b} and
      \ref{fig:2}{\bf c-d}. Results for fixed $n$ and growing $B$ from Figure \ref{fig:4} result from vertical cuts of
      the map. Other possible progressions $B=B(n)$ can be charted similarly. 

      The large-fluctuations regime is a salient anomaly (Figure \ref{fig:5}{\bf b-c}) expanding up- and right-wards
      (perhaps unboundedly) over a broad range of $(n, B)$ values. Its contour constraints dependencies, $B=B(n)$, that
      could avoid this regime. Its upper bound seems to grow faster than $n\cdot\log(n)$ suggesting that case $III$ with
      $B_0 = 50$ will not scape large-fluctuations despite sustained growth. 

      Supporting Figure \ref{fig:SI8} shows maps for $\left< h^a \right>$ and $\left< r \right>/n$. The dent of low
      $\left< h^a \right>$ due to within-game coordination in simple yet wealthy setups is notable (Supporting Figure
      \ref{fig:SI8}{\bf a}). We argued that such cartel-like cases could be defused by increasing complexity. But this map
      shows that, if $n$ grows too much without raising $B$, $\left<h^a\right>$ drops gradually -- consensus strategies
      build up again. It is intuitive that $\left< r \right> / n$ grows alongside $B$ (Supporting Figure \ref{fig:SI8}{\bf
      b}), since more available resources can be dedicated to rigging the games. Less intuitively, our map shows $\left<
      r \right> / n$ growing with $n$ as well, even if returns per game diminish. We speculate that, for low $n$,
      different agents meddling are likely to collide, resulting in uncertain returns. With larger $n$,  different
      agents can intervene different degrees of freedom, lowering the chance of mutual frustration.

  \section{Discussion} 
    \label{sec:4}

    It is difficult to pinpoint what an `unrigged' economy is. We model economies as containing degrees of freedom that
    can be controlled at a cost by its actors. Unchecked degrees favor economic agents at random. An economy with more
    `riggable' facets is more complex. We studied dynamics, stability, and viability of a rigged economy toy model as
    its complexity and total wealth change. 

    Simple scenarios allow a study of equilibria in payoff matrices. We find that individual degrees of freedom turn
    from minority into majority games, through a dynamical phase, as intervention raises. Agent-based simulations
    confirm these regimes. They also show new behaviors as synergies develop between degrees of freedom. These new
    behaviors (difficult to capture with payoff matrices) halt within-game coordination. Within-game coordination in
    simple yet wealthy markets resembles cartels: most economic actors with decision power bend the rules homogeneously
    in their favor. Our results suggest that this consensus is spontaneously defused if the system becomes complex
    enough, which can be empirically tested. 

    We study our toy economies as their complexity increases and the wealth they distribute remains constant, grows
    linearly, or faster than linearly with the number of economic degrees of freedom. In general, wealth should grow
    faster than linearly. Against raising complexity, stagnant or slowly growing wealth only sustains a decreasing
    ensemble of actors sharing ever more meager resources. An unlucky fluctuation can kill them off. This becomes more
    pressing as our model predicts that large fluctuations build up abruptly above a complexity threshold. These large
    fluctuations remind us of chaotic regimes in the El Farol and similar problems \cite{Gintis2000, Whitehead2008,
    Buchanan2008, Sornette2017}. In them, agents with sufficient rationality anticipate a market, but their own success
    turns the market unpredictable. In our model, above a complexity threshold, non-competitive intervention choices
    become intertwined across games. Birth and death of agents ripple system-wide, making successful strategies hard to
    track. Even though agents are exploring non-competitive strategies, large fluctuations ($\sim 15-20\%$ population
    size) ensue, compromising the system's viability -- thus non-competitive actions can result in negative spillover by
    sheer market complexity. This is another testable conclusion. 

    Behavioral economics offers a prominent chance to test our findings. We see stable states with raising rigging
    pressure as expected returns grow. This is consistent with empirical data on {\em cheating}: while different
    profiles exist (including people hardly corrupted), cheating eventually ensues for large enough rewards
    \cite{HilbigThielmann2017}, especially after removing the concern of being caught \cite{KajackaiteGneezy2017}.
    Further experiments reveal that cheating is more likely as a partnership \cite{WeiselShalvi2015}. This resounds with
    our model's ``cartels'' in simple yet wealthy economies. Such simple experiments are perfect to test our predictions
    for growing complexity: Does coordination fall apart swiftly? Does rigging pressure grow with complexity in the long
    run? More ambitiously, we could emulate recent implementations of Prisoner dilemmas and other simple games
    \cite{Cassar2007, KirchkampNagel2007, TraulsenMilinski2010, GrujicSanchez2010, SuriWatts2011, GraciaMoreno2012,
    GrujicSanchez2014, RandChristakis2014, MaoWatts2017, Sanchez2018}. 

    This work did not aim at specific realism, but at capturing elements that we find essential about `rigged'
    economies, and thus derive qualitative regimes and wealth-complexity scalings that keep our toy economies viable.
    Exploring lesser model parameters (Appendix \ref{app:3}), the same phenomenology features consistently. This
    encourages us to think that we are unveiling general results of `rigged' economies. But we made important
    simplifications to keep our model tractable. All agents participate of all games, while real economic actors might
    walk out or be banned from a specific market. We model all degrees of freedom with a similar game. Real
    manipulations might treat agents with a same strategy differently. Some pay off only the first intervention, others
    reward non-linearly a varying investment. Exploring these and other alternatives is easy and might uncover new
    systemic regimes. Our results constitute solid limit behaviors that should be recovered under appropriate
    circumstances. 

    In our model, wealth is generated externally -- the economic games merely distribute it. An important variation
    should create wealth organically, depending on population size, strategies explored, and degrees of freedom
    available. These, like technological niches, are developed and sustained at a cost. Rigged economies might then
    correct themselves by losing complexity if necessary. Similar feedbacks can poise complex systems near critical
    regimes \cite{BakWiesenfeld1987, BakWiesenfeld1988, BakSneppen1993, Kauffman1993, Kauffman1996, Bak1996,
    DickmanZapperi2000, Munoz2018}, which proved relevant to rationalize some phenomenology in economics
    \cite{Sornette2006, Sornette2017, LiLi2019} -- at criticality we observe fat tails in wealth distributions or
    dynamic turnover of complex markets. 

    An important design choice are the Darwinian dynamics that propagate successful strategies. We could have modeled
    boundedly rational agents that learn, similarly spreading successful behaviors. A key parameter then would be a
    learning rate, instead of our replication cost, $C_C$. Similar models show that certain regimes depend tangentially
    on the cognitive mechanism \cite{Gintis2000, Whitehead2008, Buchanan2008, Sornette2017}. Different implementations
    might move around the onset of unpredictable regimes (as $C_C$ does, Supporting Figure \ref{fig:SI8.5}). When
    unpredictability is intrinsic to the phenomena studied, rational agents cannot perform better either. Our results
    suggest that rigged economies might be intrinsically uncomputable in certain limits. 

    Our work is designed in economic terms, but it has an obvious political reading -- e.g. construction and control of
    power structures. More pragmatically, in our model wealth redistribution is achieved through low rigging pressures.
    Empirical measurements of redistribution might help us map real economies into our framework, as has been done for
    similarly abstract models \cite{DevittLeeBoghosian2018, LiLi2019}. At large, the evolutionary stability of fair
    governance \cite{Barucca2020} is under scrutiny. In ecosystems that bring together wealth, people, and economic
    games, all subjected to Darwinism: What lasting structures emerge? Do fair rules survive? Under which circumstances
    does unfairness prevail?

\vspace{0.2 cm}

  \section*{Acknowledgments}

    The author wishes to thank Roberto Enr\'iquez (Bob Pop) for his deep insights in socio-economic systems, which
    prompted this work. The author also wishes to thank Juan Fern\'andez Gracia, V\'ictor Egu\'iluz, Carlos Meli\'an,
    and other IFISC (Institute for Interdisciplinary Physics and Complex Systems) members, as well as V\'ictor Notivol,
    Paolo Barucca, David Wolpert, Justin Grana, and, very especially, Federico Curci for indispensable feedback about
    economic systems. This work has been funded by IFISC as part of the Mar\'ia de Maeztu Program for Units of
    Excellence in R\&D (MDM-2017-0711), and by the Spanish National Research Council (CSIC) and the Spanish Department
    for Science and Innovation (MICINN) through a Juan de la Cierva Fellowship (IJC2018-036694-I).

\newpage
\appendix 

  \setcounter{figure}{0}
  \setcounter{table}{0}
  \setcounter{page}{1}
  \renewcommand{\figurename}{SUP. FIG.}

  \section{Payoff matrices for simple cases} 
    \label{app:1} 

    Let us note that the model is actually grounded on game theory by building payoff matrices for simple scenarios.

    \begin{table*}
      \begin{center}
        \begin{tabular}{/c/c|c/c|c/}
          \Xhline{4\arrayrulewidth}
          Player's behavior & $a=0$, $r=0$ & $a=1$, $r=0$ & $a=0$, $r=1$ & $a=1$, $r=1$ \\ 
          \hline
          Payoff & $b/2$ & $b/2$ & $b - C_R$ & $b -C_R$ \\
          \Xhline{4\arrayrulewidth}
        \end{tabular}

        \caption{{\bf Payoff matrix for one game with one player.} The agent's behavior is coded by two bits. A first
one ($a$) indicates the agent's strategy ($0$ or $1$). A second bit ($r$) indicates whether the agent attempts to rig
the game or not. }
        
        \label{tab:ST1}
      \end{center}
    \end{table*}

    Take one game ($n=1$) and a fixed population of one player $N(t)=1$ (i.e. even if the agent accumulates wealth, she
    does not have descendants, so she never pays $C_C$; she is not removed either if she accumulates negative wealth).
    Supporting Table \ref{tab:ST1} shows the average payoff that a player earns if she plays the same game repeatedly
    with fixed behavior (i.e. fixed strategy $a^1_1$ and rigging choice $r^1_1$). If there is no intervention
    ($r^1_1=0$), the winning rule $R(t)$ is set randomly at each iteration and the expected payoff per round is $b/2$.
    If the agent attempts to rig the game ($r^1_1=1$), she always succeeds (because there is no opposition) and sets
    $R(t)$ equals to its own strategy ($R(t) = a_1$). Thus she ensures earning an average $b$ per round, from which
    $C_R$ must be subtracted. The optimal strategy is to intervene if $C_R < b/2$. 

    Things become more interesting if we add another agent ($N(t)=2$) while, again, playing only one game. This case was
    discussed in the main text. Let us take a closer look. There are three scenarios worth considering separately, and
    each corresponds to a $2 \times 2$ block matrix from Table \ref{tab:1}: 
      \begin{itemize}

        \item {\bf No player attempts any rigging} (upper-left block matrix in Table \ref{tab:1}). In this case the
winning rule is set randomly, so that both players win half of the time. If they play the same strategy, whenever they
win (i.e. half of the rounds), they must split the earnings. If they play different strategies, each agent still wins
half of the time but they always get to keep all the earnings. In other words, in this case the model reduces to a
minority game. If the winning rule behaves randomly because there is no intervention, the preferable strategy is to stay
in the minority. This is true also when there are more players (see below), since the only varying factor that reduces a
player's earnings is the number of others with a same strategy, among whom the benefit is split.

        \item {\bf Only one of the players attempts to rig the game} (either off-diagonal block matrices in Table
\ref{tab:1}). The intervening agent pays $C_R$ to ensure that the winning strategy $R(t)$ always matches her own.
Assuming that only one agent (e.g. agent $1$, thus look at the top-right block matrix in Table \ref{tab:1}) is given the
option to rig the game, doing it becomes always favorable if $C_R < b/4$, disregarding of what action agent $2$ takes.
If $b/4 < C_R < b/2$, then rigging the game is favorable only if agent $2$ plays a different strategy. If $C_R > b/2$,
it never becomes favorable to rig the game. 

        \item {\bf Both agents attempt to rig the game} (bottom-right block matrix in Table \ref{tab:1}). Both agents
pay $C_R$ in this case. But they only intervene the game successfully if both play the same strategy. Note that this has
the effect of turning the minority game into a neutral one regarding the agent's strategies $a^i_1$: If both players are
attempting to intervene the game, they will always receive the same payoff disregarding of whether $a^1_1=a^2_1$ or not.
The received payoff is always less than the best scenario with no intervention. But, if $C_R < b/4$, it is better than
the scenarios with no intervention and matching strategies. 

      \end{itemize}

    It becomes cumbersome to write payoff matrices when more players are involved, but it is still feasible for
    $N(t)=3$. We do so in Supporting Tables \ref{tab:ST3}, \ref{tab:ST4}, and \ref{tab:ST5}. In them, we group up the
    behaviors of players $2$ and $3$, assuming that, whenever one of the three agents plays a different strategy (i.e.
    not all $a^i_1$ are the same), it is always player $1$ (either $a^1_1=0$ and $a^{2,3}_1=1$ or $a^1_1=1$ and
    $a^{2,3}_1=0$). We call player $1$ the minority player and players $2$ and $3$ the majority players. 

    \begin{table*}[t]
      \begin{center}
        \includegraphics[width=0.8\textwidth]{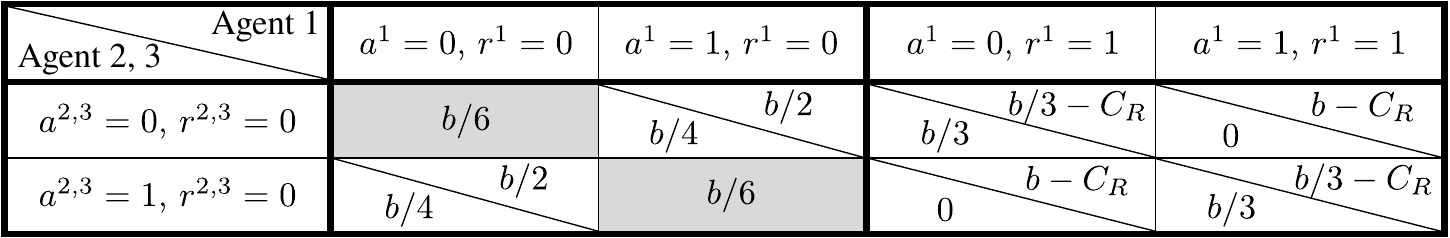}

        \caption{{\bf Payoff matrix of one game with three players -- only the minority player can rig. } We assume that
player $1$ is in the minority when there is no consensus. In this table, only player $1$ is allowed to rig the game, so
she always succeeds. Entries marked in gray are global Nash equilibria when rigging is very expensive $C_R >> b$. }

        \label{tab:ST3}
      \end{center}
    \end{table*}

    \begin{table*}[t]
      \begin{center}
        \includegraphics[width=0.8\textwidth]{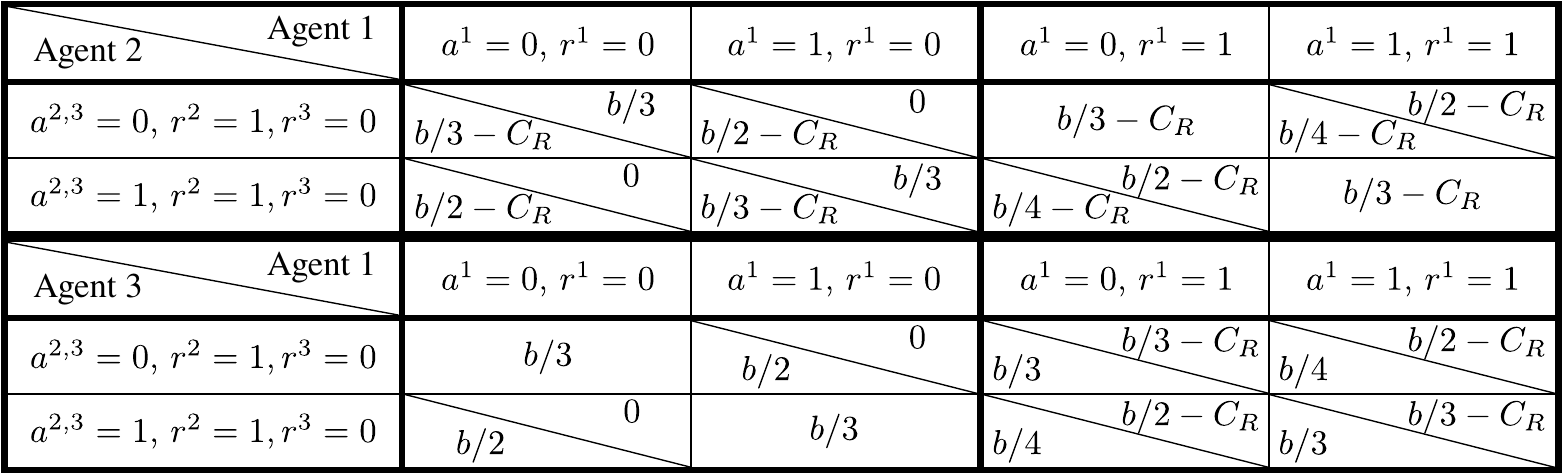}

        \caption{{\bf Payoff matrix of one game with three players -- only one of the majority players (player $2$)
rigs. } This is the only situation in which the symmetry between the majority players is broken. }

        \label{tab:ST4}
      \end{center}
    \end{table*}

    \begin{table*}[t]
      \begin{center}
        \includegraphics[width=0.8\textwidth]{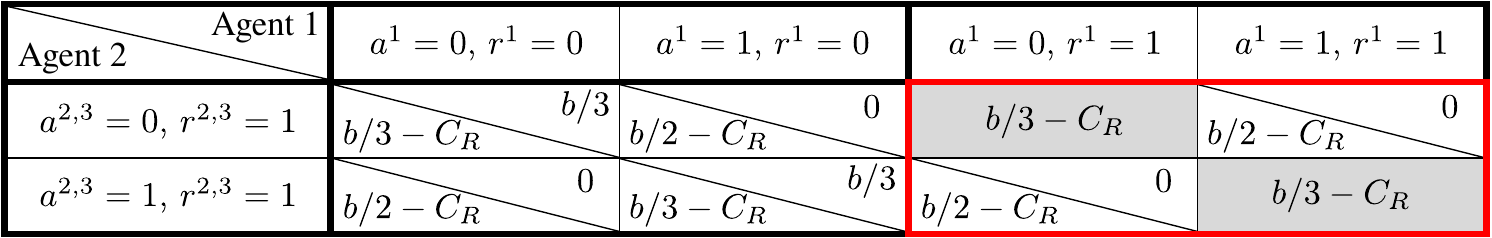}

        \caption{{\bf Payoff matrix of one game with three players -- both majority players rig the game. } Since they
are in the majority, they always succeed in their attempt to set the winning rule. If all three players rig the game
simultaneously (red frame), the model turns into a majority game -- i.e. the best strategy is to play what everybody
else is playing. Gray squares indicate Nash equilibria of this sub-game. }

        \label{tab:ST5}
      \end{center}
    \end{table*}

    Supporting Table \ref{tab:ST3} shows the average payoff matrix when only the minority player is allowed to rig the
    game. If she is not meddling with the rules (left half of Supporting Table \ref{tab:ST3}) we deal again with a
    minority game. The Nash equilibria of this subgame ($r^i_1=0 \> \forall i$ and $a^1_1 \ne a^{2,3}_1$) are Nash
    equilibria of the whole game if $C_R > b/4$. For cheaper cost of rigging, the global Nash equilibria disappear as it
    becomes favorable to one of the majority agents to intervene (for which we have to look at Supporting Table
    \ref{tab:ST4}). This sets on a dynamic situation similar to the one discussed in the main text.  Finally, Supporting
    Table \ref{tab:ST5} has both majority players attempting (and succeeding, since they are in the majority) to rig the
    game. Interestingly, if all three agents are trying to manipulate the winning rule (red frame), the model turns into
    a majority game in which all three players earn $b/3 - C_R$. If  an agent decided to change its strategy $a^i_1$,
    this would put her in the minority, in which (according to Supporting Table \ref{tab:ST5}), it would earn $0$ per
    round -- thus full coordination is a Nash equilibrium of the subgame in which everybody intervenes. This, however,
    is not a Nash equilibrium of the complete game: in full coordination, it would pay off to a single agent to stop
    rigging the game. This suggests that the way that our model reaches large levels of coordination (as discussed in
    the main text) is a tragedy-of-the-commons scenario. 

    In a static situation (i.e. population is fixed and agents always choose the same actions and whether to rig each
    game or not), games are independent of each other. We could take these payoff matrices and compute averages over
    many games. The situation becomes more difficult when dynamics are included. Because new agents can be born and
    older ones may die, averages over time should keep into account that agent's actions feed back on each other. For
    example, a possible good strategy for agent $1$ may be to rig games that favor a third agent (say, agent $3$) who,
    in turn, is rigging games that favor agent $1$. In such a way, games can become coupled to each other and result in
    much more complicated payoff functions.

  \section{Supporting plots and discussion for increasing economy size and fixed complexity}
    \label{app:2}

    Despite its simplicity, the model turned out to have very rich dynamics. Its behavior changes, sometimes
    drastically, with the economy complexity (as measured by the number of games, $n$) or with its size (as measured by
    wealth distributed at each round, $B$). In this appendix we take a closer look to what happens when the number of
    games is fixed, but the wealth distributed in each game grows. We saw an example of this (with $n=2$) in the main
    text, and we saw that increasing the number of resources drives agents to coordinate with each other in different
    manners around the available strategies and whether to rig them or not. 

    \begin{figure*}
      \begin{center}
        \includegraphics[width=\textwidth]{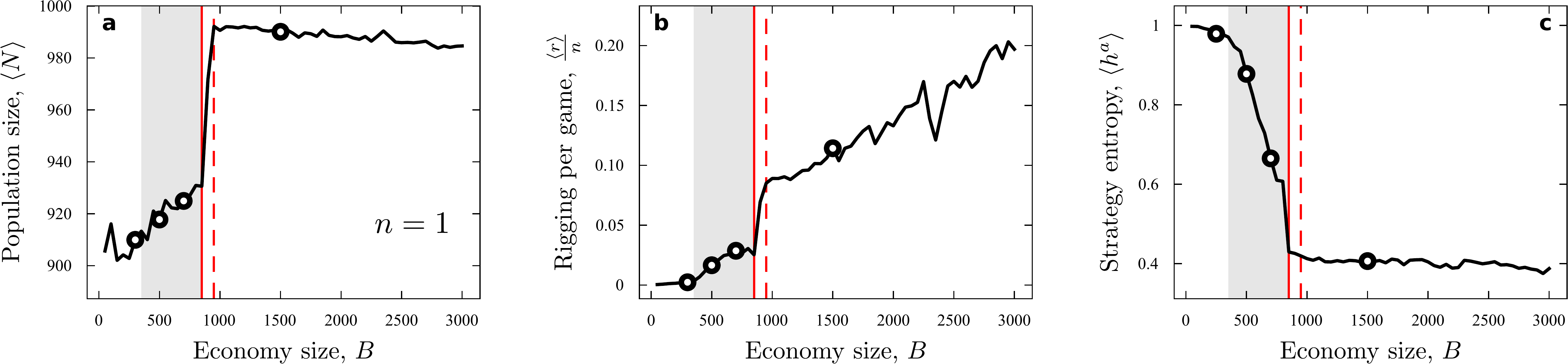}

        \caption{{\bf Fixed economy complexity, $n=1$, and growing distributed wealth, $B$.} {\bf a} Average population
size in the steady state. {\bf b} Rigging pressure over the only game in the steady state. {\bf c} Strategy entropy in
the only game in the steady state. Circles over the plots curves indicate values of $B$ for which we show samples of the
dynamics in Supporting Figure \ref{fig:SI2}. }

        \label{fig:SI1}
      \end{center}
    \end{figure*}

\begin{sidewaysfigure*}
      \begin{center}
        \includegraphics[width=0.8\textwidth]{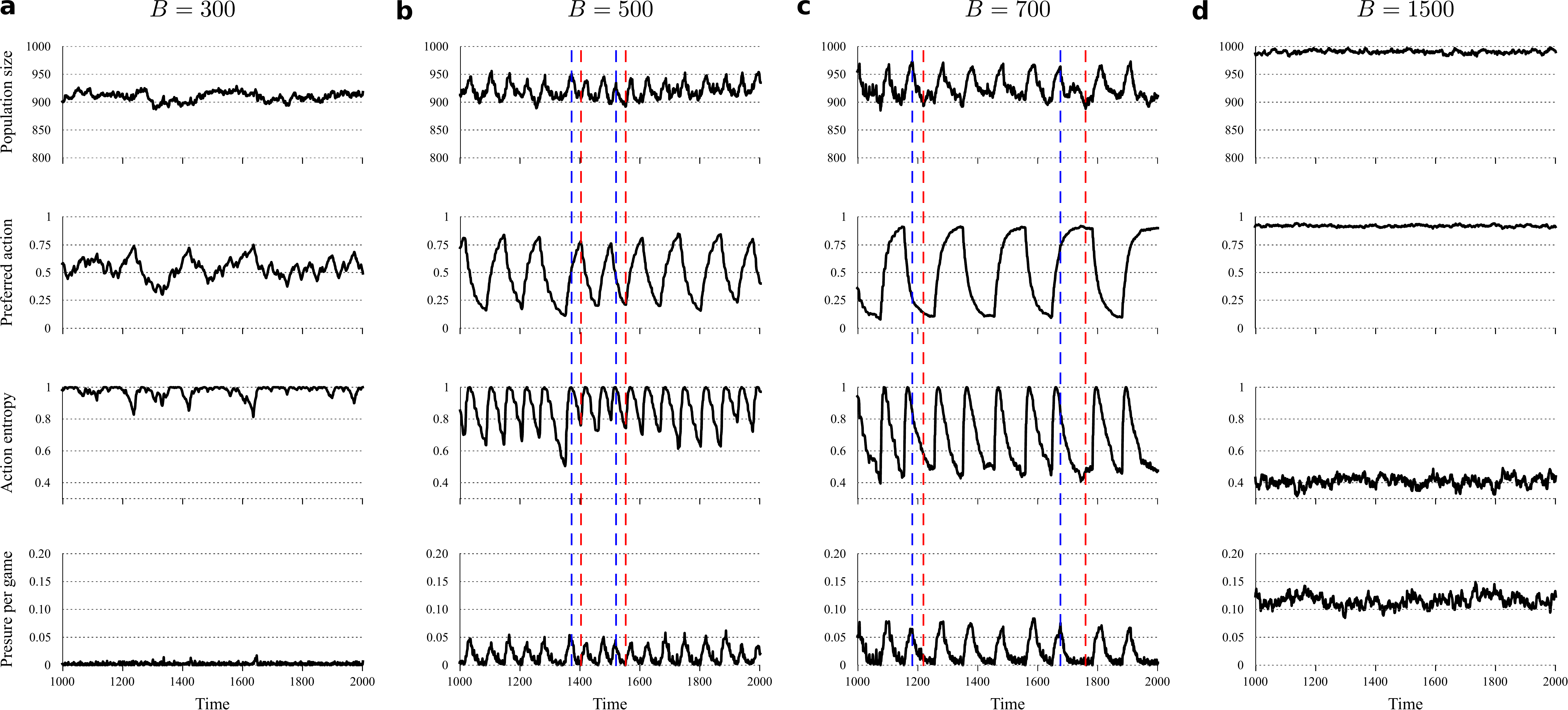}

        \caption{{\bf Samples of the model dynamics for fixed economy complexity ($n=1$) and different economy sizes.}
The model dynamics over $1000$ iterations are shown for one game and economy sizes $B=300$ ({\bf a}), $B=500$ ({\bf b}),
$B=700$ ({\bf c}), and $B=1500$ ({\bf d}). Panels in the top row show the population size over time, second row shows
the fraction $f_1(t)$ of agents playing strategy $1$, third row shows the action entropy ($h^a = -f_1\cdot log_2(f_1) -
(1-f_1)\cdot \log_2(1-f_1)$) which is maximal when the agents disagree maximally over their strategies regarding the one
game, bottom row shows the rigging pressure over time. }

        \label{fig:SI2}
      \end{center}
    \end{sidewaysfigure*}

    Let us start with the simplest case now, with $n=1$. We plot average population size in the steady state (Supporting
    Figure \ref{fig:SI1}{\bf a}), rigging pressure over the only game (Supporting figure \ref{fig:SI1}{\bf b}), and the
    strategy entropy (Supporting Figure \ref{fig:SI1}{\bf c}) as more resources become available. Circles over these
    plots show values of $B$ for which we show $1000$ iterations of the dynamics in Supporting Figure \ref{fig:SI2}. 

    As we saw for $n=2$ in the main text, if there are very few resources, spending them in intervening the economy is
    not a favored behavior in the steady state. This implies that the winning strategy is randomly $0$ or $1$, likely
    changing from one iteration to the next. As a consequence, the steady population does not settle for either
    strategy. The second row of Supporting Figure \ref{fig:SI2}{\bf a} shows $f_k$ the fraction of agents playing $1$
    over time in game $k=1$. We see that this number moves around $0.5$, indicating that roughly half the population is
    choosing $1$ and the other half is choosing $0$. This results in a high strategy entropy $\left<h^a_k\right>$ in
    game $k=1$, as shown in the third row of Supporting Figure \ref{fig:SI2}{\bf a}. The fourth row shows that, indeed,
    the rigging pressure is negligible for low values of $B$. 

    Above some amount of available resources, an effective level of rigging pressure starts to build up periodically
    (lower row of Supporting Figure \ref{fig:SI2}{\bf b-c}). Conceive a situation in which no rigging exists, but a
    mutation produces a single agent that decides to rig the game. She secures the next rounds played, and all agents
    playing her same strategy are consequently benefited. She will replicate, producing more agents that rig the game in
    the same direction. But those {\em parasitic} agents playing her same strategy without paying $C_R$ will earn more
    and produce a slightly larger descent. Darwinian dynamics expand slight differences exponentially over time, thus
    eventually the rigging agents are driven off to extinction. Some agents playing the other option might have survived
    -- perhaps because they had some savings. After all rigging agents were removed from the population, those playing
    the minority option will earn more money (as our analysis of payoff matrices indicates), and start making a come
    back. Eventually, a mutation in their descendants might produce an agent rigging the game to favor the minority.
    This would start a cycle all over again. Supporting Figure \ref{fig:SI2}{\bf b-c} shows that this oscillating
    behavior takes place for a range of economy sizes $B$. 

    Vertical, dashed blue lines in this figure show the point at which population reaches a maximum, which happens as
    the rigging pressure peaks as well. Interestingly, this is also the point at which the proportion of agents playing
    either strategy is well balanced. The population minimum (indicated by vertical, dashed red lines) happens when the
    rigging pressure is minimal as well and the population presents a more homogeneous strategy. Increasing the economy
    size results in longer alternating cycles of this nature. Perhaps, hoarding more resources might make it more
    difficult to remove agents rigging the games in one direction over the other. 

    Eventually, these cycles become infinitely long so that most of the population ends up adopting a same strategy
    (Supporting Figure \ref{fig:SI2}{\bf d}). This homogeneous state supports a larger population than the cycling or
    fully random regimes. However, convergence to the majority is not full. A reservoir of agents playing the minority
    strategy survives, suggesting that these can occasionally succeed in rigging the game and upset the majority. \\ 

    \begin{figure*}
      \begin{center}
        \includegraphics[width=\textwidth]{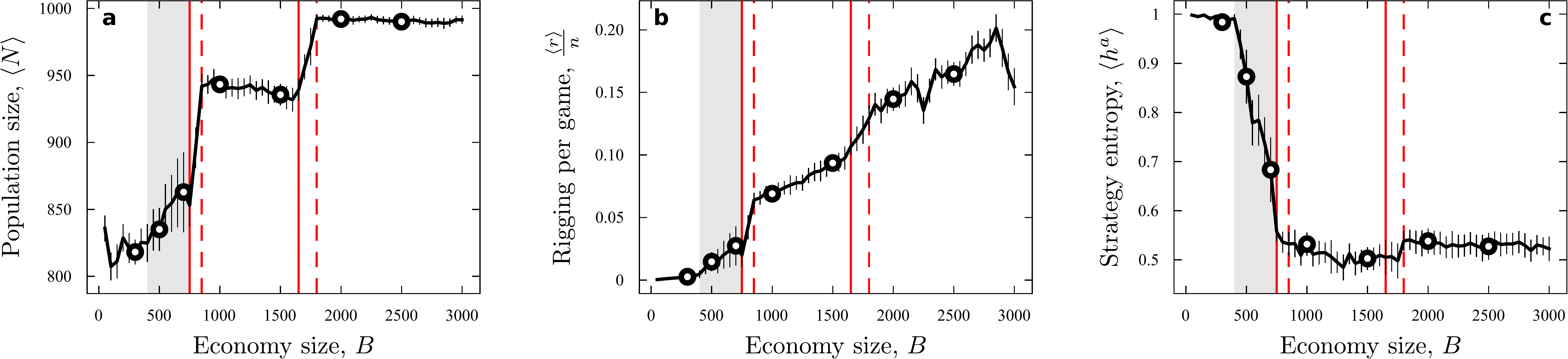}

        \caption{{\bf Fixed economy complexity, $n=2$, and growing distributed wealth, $B$.} Average population size in
the steady state. {\bf b} Average rigging pressure per game. {\bf c} Average strategy entropy over the two games.
Circles over the plots curves indicate values of $B$ for which we show samples of the dynamics in Supporting Figure
\ref{fig:SI4}. Error bars indicate the standard deviation over the last $500$ iterations of the corresponding
simulation. }

        \label{fig:SI3}
      \end{center}
    \end{figure*}

\begin{figure*}
      \begin{center}
        \includegraphics[width=\textwidth]{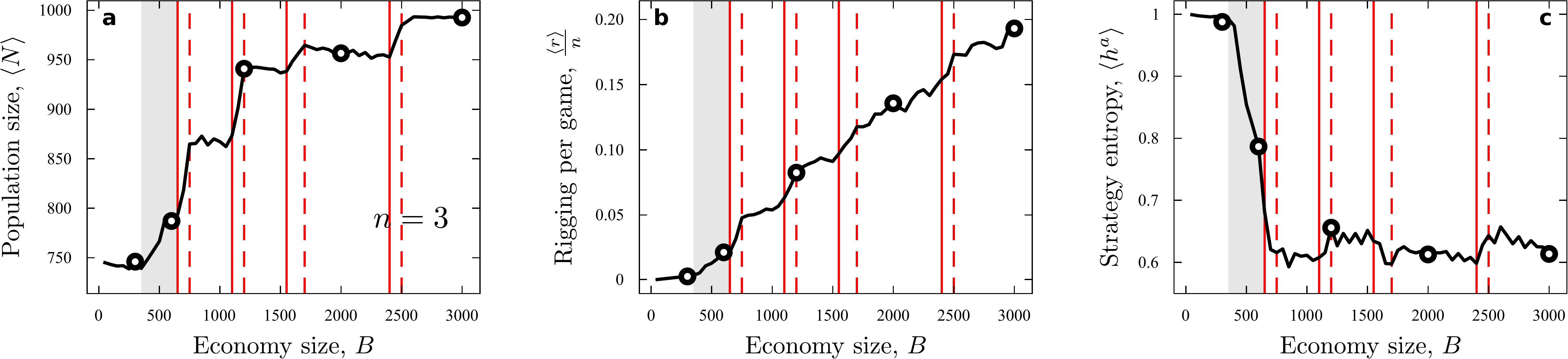}

        \caption{{\bf Fixed economy complexity, $n=3$, and growing distributed wealth, $B$.} Average population size in
the steady state. {\bf b} Average rigging pressure per game. {\bf c} Average strategy entropy over the three games.
Circles over the plots curves indicate values of $B$ for which we show samples of the dynamics in Supporting Figure
\ref{fig:SI6}. }

        \label{fig:SI5}
      \end{center}
    \end{figure*}

\begin{figure*}
      \begin{center}
        \includegraphics[width=\textwidth]{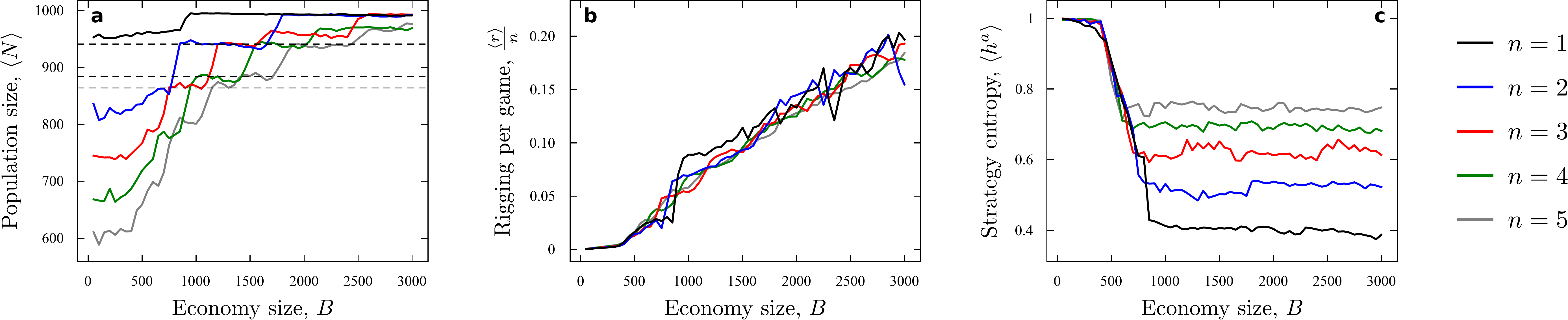}

        \caption{{\bf Fixed economy complexity, $n$, and growing distributed wealth, $B$ -- comparison of cases with
small complexity.} {\bf a} Average population size in the steady state. Horizontal dashed lines help us identify cases which,
despite having different $n$ and $B$, reach similar stable population size. This suggests that some scenarios might be
essentially equivalent. {\bf b} Average rigging pressure per game. {\bf c} Average strategy entropy over the $n$ games
in each case. }

        \label{fig:SI6.5}
      \end{center}
    \end{figure*}

\begin{figure*}
      \begin{center}
        \includegraphics[width=\textwidth]{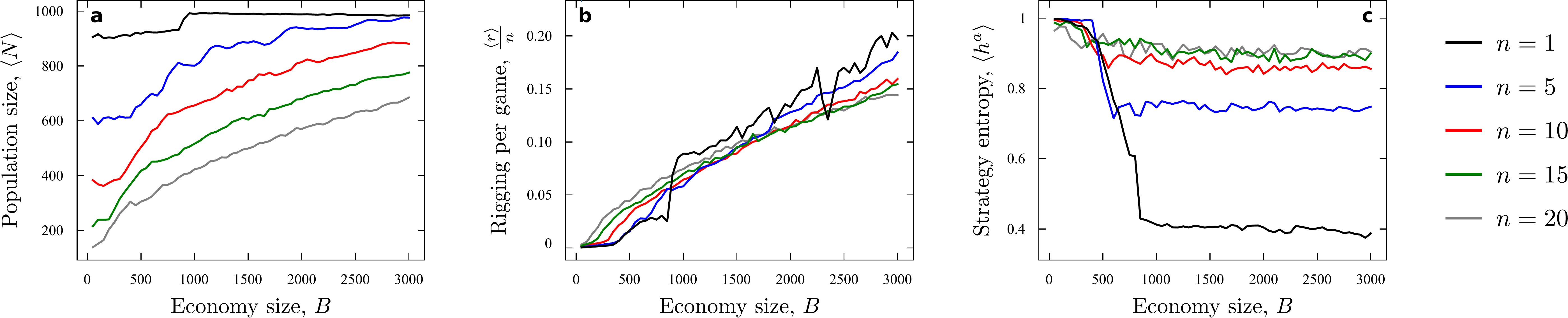}

        \caption{{\bf Fixed economy complexity, $n$, and growing distributed wealth, $B$ -- comparison of cases with
large complexity.} {\bf a} Average population size in the steady state. Boost in stable population size are smoothed
into a continuous buildup as complexity increases. {\bf b} Average rigging pressure per game. {\bf c} Average strategy
entropy over the $n$ games in each case. }

        \label{fig:SI7}
      \end{center}
    \end{figure*}

\begin{sidewaysfigure*}
      \begin{center}
        \includegraphics[width=\textwidth]{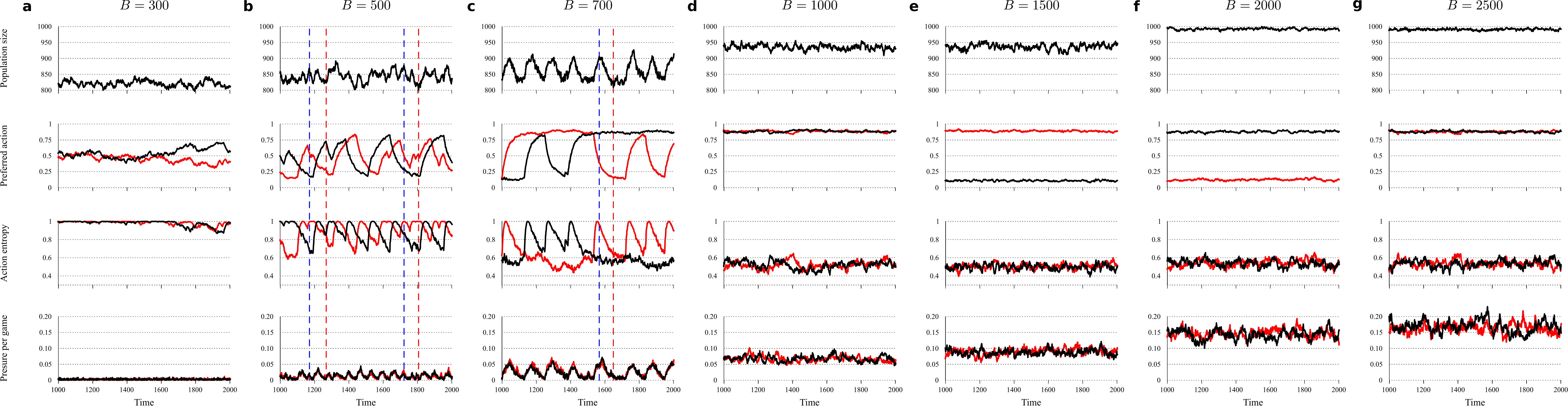}

        \caption{{\bf Samples of the model dynamics for fixed economy complexity ($n=2$) and different economy sizes.}
The model dynamics over $1000$ iterations are shown for one game and economy sizes $B=300$ ({\bf a}), $B=500$ ({\bf b}),
$B=700$ ({\bf c}), $B=1000$ ({\bf d}), $B=1500$ ({\bf e}), $B=2000$ ({\bf f}), and $B=2500$ ({\bf g}). Panels in the top
row show the population size over time, second row shows the fraction $f_k(t)$ of agents playing strategy $1$ in the
$k$-th game, third row shows the corresponding action entropy ($h^a = -f_k\cdot log_2(f_k) - (1-f_k)\cdot
\log_2(1-f_k)$) which is maximal when the agents disagree maximally over their strategies regarding the one game, bottom
row shows the rigging pressure over time. }

        \label{fig:SI4}
      \end{center}
    \end{sidewaysfigure*}

    Regarding $n=2$, in Supporting Figure \ref{fig:SI3} we reproduce the first three panels of Figure \ref{fig:4}. Here
    we have added error bars (which indicate the standard deviation of each quantity over the last $500$ iterations of
    each simulation) to give an idea of the variation that we can find. Error bars are of similar relative magnitude in
    all examples shown (Supporting Figure \ref{fig:SI1} for $n=1$ and Supporting Figure \ref{fig:SI5} for $n=3$, as well
    as Supporting Figures \ref{fig:SI6.5} and \ref{fig:SI7} which compare several $n$). Hence, we omitted error bars
    anywhere else for clarity. Figure \ref{fig:4} and Supporting Figure \ref{fig:SI3} both show a transition (similar to
    the one observed for $n=1$) from unintervened games (sample dynamics are shown in Supporting Figure
    \ref{fig:SI4}{\bf a}), through cycles of growing and declining coordination (Supporting figure \ref{fig:SI4}{\bf
    b-c}), to more homogeneous states (Supporting Figure \ref{fig:SI4}{\bf d-g}). 

    In the main text we indicated that this cyclic regime (which we called a shift towards within-game coordination) is
    transited simultaneously for both games. There are some nuances, though. Supporting Figures \ref{fig:SI4}{\bf b-c}
    show that for both cases the amount of rigging in both games is very similar (bottom row). But $\left<h^a_k\right>$
    reveals important asymmetries which, furthermore, change as we increase $B$. For the lowest $B$ value shown with
    cyclic behavior ($B = 500$, Supporting Figures \ref{fig:SI4}{\bf b}), in average, the population does not converge
    on persistent homogeneous strategies for neither of the games. But it does not stay divided randomly either (as it
    happens for $B=300$, Supporting Figure \ref{fig:SI4}{\bf a}). In the second example with oscillating behavior ($B =
    700$, Supporting Figure \ref{fig:SI4}{\bf c}), in average, the population has converged regarding the strategy of
    one of the games. The dynamics move towards converge for the other game as well, but they fail periodically or, if
    they succeed, then the strategy for the other game (formerly homogeneous) breaks apart. Summing up: while the level
    of intervention is similar in both games, this symmetry is broken regarding how homogeneous the population is about
    each strategy. 

    After the cyclic behavior, Figure \ref{fig:4} and Supporting Figure \ref{fig:SI3} still show two more regimes
    separated each by a large boost in stable population size. As noted in the main text, this last regime shift is not
    accompanied by large changes in action entropy, $\left<h^a_k\right>$, of neither game. Hence, we conclude that
    within-game coordination has been exhausted and that new kinds of correlations, now across games, are taking place.
    Supporting Figure \ref{fig:SI4}{\bf d-g} are samples of the dynamics in those two regimes after the cyclic behavior.
    Panels \ref{fig:SI4}{\bf d-e} sample the regime between $B \sim 800$ and $1600$, and panels \ref{fig:SI4}{\bf f-g}
    sample the regime for $B > 1700$. The second row shows that in all four cases the strategies have become homogeneous
    across the population for both games -- still with reservoirs of agents playing a minority strategy. The level of
    homogeneity remains roughly the same for both regimes -- indicated by $\left<h^a_k\right>$ as well, which remains
    low throughout. We appreciate the population boost already discussed in the main text (average population in the top
    panels of Supporting Figure \ref{fig:SI4}{\bf d-e} is lower than in Supporting Figure \ref{fig:SI4}{\bf f-g}). We
    also appreciate that the fluctuations in population is larger when the population is smaller. The other significant
    difference between these two last regimes is that the rigging pressure becomes notably higher in the regime with
    larger $B$. This suggests that the difference between both regimes lies in a more efficient coordination between the
    rigging efforts across games, allowing the population to extract more wealth in average. For example, we see that
    both games sustain a minority of agents playing the minority option even if the population has broadly converged
    about each game's strategy. But these minority-playing agents might not be the same in both games if the conditions
    allow it. A transition to a higher across-game coordination might happen if the agents playing the minority in both
    games become the same. \\

    Finally, for $n=3$ too, we show average population size in the steady state (Supporting Figure \ref{fig:SI5}{\bf
    a}), rigging pressure over each game (Supporting Figure \ref{fig:SI5}{\bf b}), and action entropy (Supporting Figure
    \ref{fig:SI5}{\bf c}) as more resources become available (i.e. as $B$ grows). We observe more regime shifts (as
    identified by boosts in population size) than for $n=2$, which is compatible with more available games and more
    possibilities for across-game coordination. 

\begin{sidewaysfigure*}
      \begin{center}
        \includegraphics[width=\textwidth]{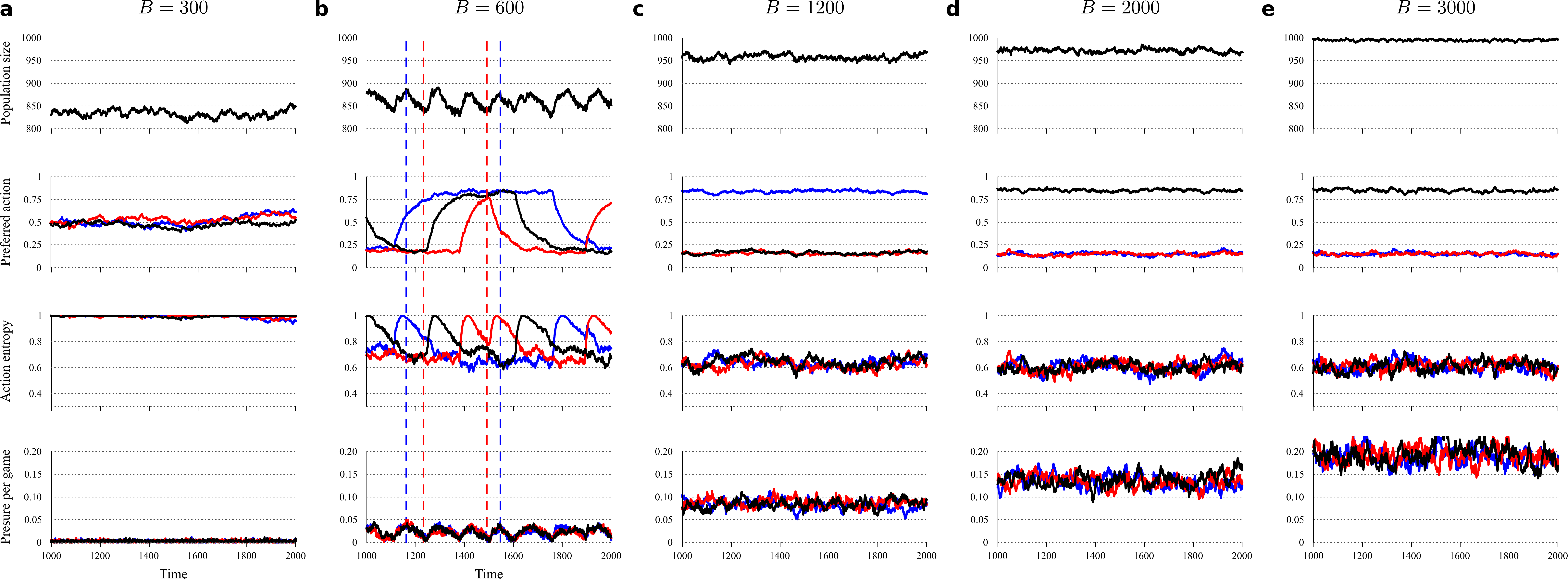}

        \caption{{\bf Samples of the model dynamics for fixed economy complexity ($n=2$) and different economy sizes.}
The model dynamics over $1000$ iterations are shown for one game and economy sizes $B=300$ ({\bf a}), $B=600$ ({\bf b}),
$B=1200$ ({\bf c}), $B=2000$ ({\bf d}), and $B=3000$ ({\bf e}). Panels in the top row show the population size over
time, second row shows the fraction $f_k(t)$ of agents playing strategy $1$ in the $k$-th game, third row shows the
corresponding action entropy ($h^a = -f_k\cdot log_2(f_k) - (1-f_k)\cdot \log_2(1-f_k)$) which is maximal when the
agents disagree maximally over their strategies regarding the one game, bottom row shows the rigging pressure over time.
}

        \label{fig:SI6}
      \end{center}
    \end{sidewaysfigure*}

    Supporting Figure \ref{fig:SI6} shows samples of the dynamics for various of these regimes, including the
    oscillatory regime. We see again that the rigging pressure over all three games is simultaneous, even if the
    symmetry regarding strategy coordination is broken. In the example shown we see that, in average, the population has
    converged regarding the strategies of two out of three games. As for $n=2$, when the population converges also about
    the third game, one of the former agreements breaks apart. There are also values of $B$ for which there is
    convergence of, at most, one game in average (not shown). Again, the extra regime shifts (which we identify by
    abrupt boosts of $\left<N\right>$) happen after within-game coordination has been exhausted. This is consistent with
    the idea that more games bring in more possible across-game coordination, which are explored only for large values
    of $B$. \\ 

    As briefly discussed in the main text, adding more games has two different effects: One the one hand more regimes
    seem to become available (as duly noted); and on the other hand, more consecutive regimes seem to be visited within
    a smaller range of $B$. This means that, as we increase $B$, regimes progress more rapidly into each-other
    (Supporting Figure \ref{fig:SI6.5}). This effect is exaggerated if even more games are available (Supporting Figure
    \ref{fig:SI7}). So much so that, instead of regime shifts, we approximate a continuous progression. The increase in
    rigging pressure per game becomes parsimonious as well (while for a small number of games it presented some discrete
    boosts associated to regime shifts). 

    Supporting Figure \ref{fig:SI6.5}{\bf c} shows that exhausting the within-game coordination results in a drop of
    strategy entropy, $\left<h^a\right>$. We see that this drop becomes less accentuated for larger $n$ (Supporting
    Figure \ref{fig:SI7}{\bf c}). This suggests that the onset of interactions across games somehow thwarts within game
    coordination. In other words, a more complex economy seems to enable populations with more diverse strategies within
    single games. 

  \section{Robustness of results against variations of model parameters} 
    \label{app:3}

    \begin{figure}
      \begin{center}
        \includegraphics[width=\columnwidth]{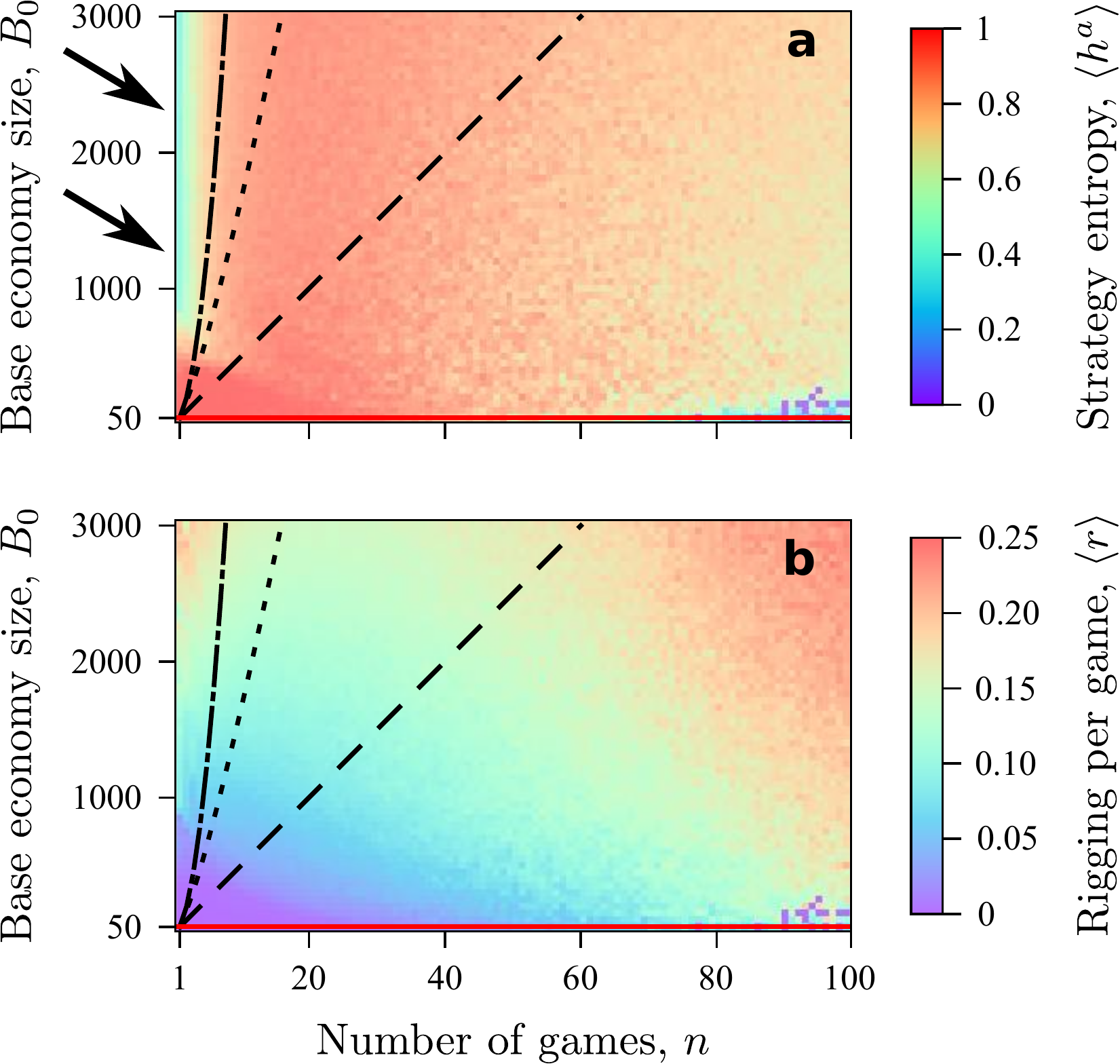}

        \caption{{\bf Comprehensive maps of strategy entropy and rigging pressure. } {\bf a} Average strategy entropy,
$\left<h^a(t)\right>$, shows a dent for small $n$ and large $B$ -- a situation that we compared to cartels in the main
text. This coordination regime falls apart swiftly as complexity grows a little bit. If we move to very large $n$
without increasing $B$, coordination starts to build up again -- yet very smoothly. {\bf b} Rigging pressure per game,
$\left<r(t)/n\right>$. Unsurprisingly, it grows with the amount of wealth distributed. More interestingly, it also grows
with the number of degrees of freedom in the system. }

        \label{fig:SI8}
      \end{center}
    \end{figure}

    The maps that we develop allow us to chart our model easily. Similarly to the maps built for $\left< N \right>$ and
    $\sigma(N) / \left< N \right>$ in Figure \ref{fig:5} of the main text, it is possible to build maps for other
    quantities such as the average strategy entropy, $\left< h^a \right>$ (Supporting Figure \ref{fig:SI8}{\bf a}); or
    the rigging pressure per game, $\left< r \right> / n$ (Supporting Figure \ref{fig:SI8}{\bf b}). Such maps can help
    us reveal regimes and phenomenology in arbitrary measurements but: How general are these phenomena? Do they depend
    critically on model parameters? 

    The model has $6$ parameters: One sets the wealth distributed by the economy in each round ($b$, $B$, and $B_0$ are
    univocally linked in each case); another one sets the available number of degrees of freedom and thus the economy's
    complexity (i.e. the number of games, $n$); two parameters set up costs (of attempting to rig a game, $C_R$; and of
    producing descent, $C_C$); a mutation rate $p_\mu$; and the parameter $N^{max}$ that sets an external upper limit to
    the population (this acts similarly to a load capacity in ecological models). We designed our model with the hope of
    pinning down essential features of rigged economies. We hope that the elements involved in the model introduce as
    few additional effects as possible. In that sense, an abundance of parameters is not desired. Furthermore, we hoped
    that the most interesting phenomenology would depend on $B$ and $n$. These parameters capture respectively the
    economy's size (as measured by distributed wealth) and complexity, which are at the center of our research
    questions. Luckily, as we show in this appendix, the observed phenomenology is not much altered when toying with the
    remaining parameters. This suggests that the regimes and phenomenology discovered for varying economy size and
    complexity should be found over again for a range of model options -- which speaks strongly in favor of the
    minimalism of our approach. 

    First we note that the cost of rigging a game $C_R$ sets a scale with respect to the wealth allotted to each game $b
    = B/n$. In our simulations we set $C_R = 1$. If we would try a different value of $C_R$, we could normalize
    $\tilde{b} \equiv b/C_R$, $\tilde{C}_C \equiv C_C/C_R$, and $\tilde{C}_R \equiv C_R/C_R = 1$ and map the parameter
    choice back to a case that we have already studied. Thus actually our model has only $5$ free parameters. \\

    \begin{figure*}
      \begin{center}
        \includegraphics[width=\textwidth]{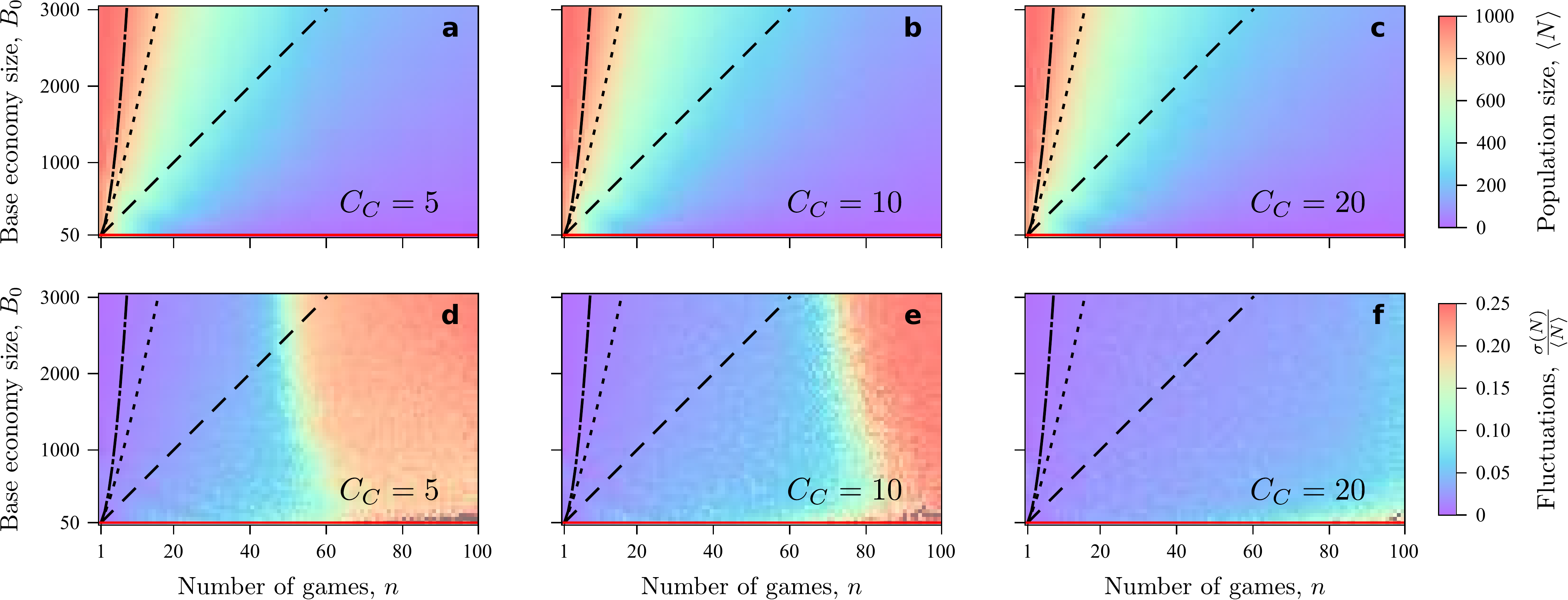}

        \caption{{\bf Comprehensive maps of $\left< N \right>$ and $\sigma(N) / \left< N \right>$ for varying
replication cost, $C_C$. } Maps result from simulating case $I$ (i.e. $B_I = B_0$, $b_I = B_0/n$) for ranges of economic
complexity, $n$, and distributed wealth, $B_0$. Black curves represent trajectories $B = B(B_0=50, n)$ for cases $II$,
$III$, and $IV$ (dashed, dotted, and dot-dashed respectively). Horizontal red lines at the bottom of each map represent
case $I$ with $B_0 = 50$. {\bf a-c} shows $\left< N \right>$ and {\bf d-f} show $\sigma(N) / \left< N \right>$. {\bf a},
{\bf d}, $C_C=5$; {\bf b}, {\bf e}, $C_C=10$; {\bf c}, {\bf f}, $C_C=20$. }

        \label{fig:SI8.5}
      \end{center}
    \end{figure*}

    Supporting Figure \ref{fig:SI8.5} shows what happens to $\left<N\right>$ and $\sigma(N)/\left<N\right>$ as $C_C$
    changes. Interestingly, the effect in $\left<N\right>$ seems negligible for the values explored (Figure Supporting
    Figure \ref{fig:SI8.5}{\bf a-c}). More notably, this parameter has the effect of displacing the onset of the
    large-fluctuations regime (Supporting Figure \ref{fig:SI8.5}{\bf d-f}). If $C_C$ is smaller, this regime ensues for
    a lower economy complexity, $n$. The parameter $C_C$ tells us how cheap it is to have descendants. When it is
    cheaper, it is easier to trigger large fluctuations; probably because a large descent explores more behaviors (both
    in rigging decisions and game strategies) simultaneously, as well as it displaces a bigger proportion of former
    agents. Both these actions result in major disruptions of the winning rules. Thus, cheaper descent more easily
    brings up a scenario in which agents are continuously deceiving each other into bankruptcy. If this is correct,
    other parameters that promote behavior diversity or population renewal among agents should have a similar effect. 

    \begin{figure*}
      \begin{center}
        \includegraphics[width=\textwidth]{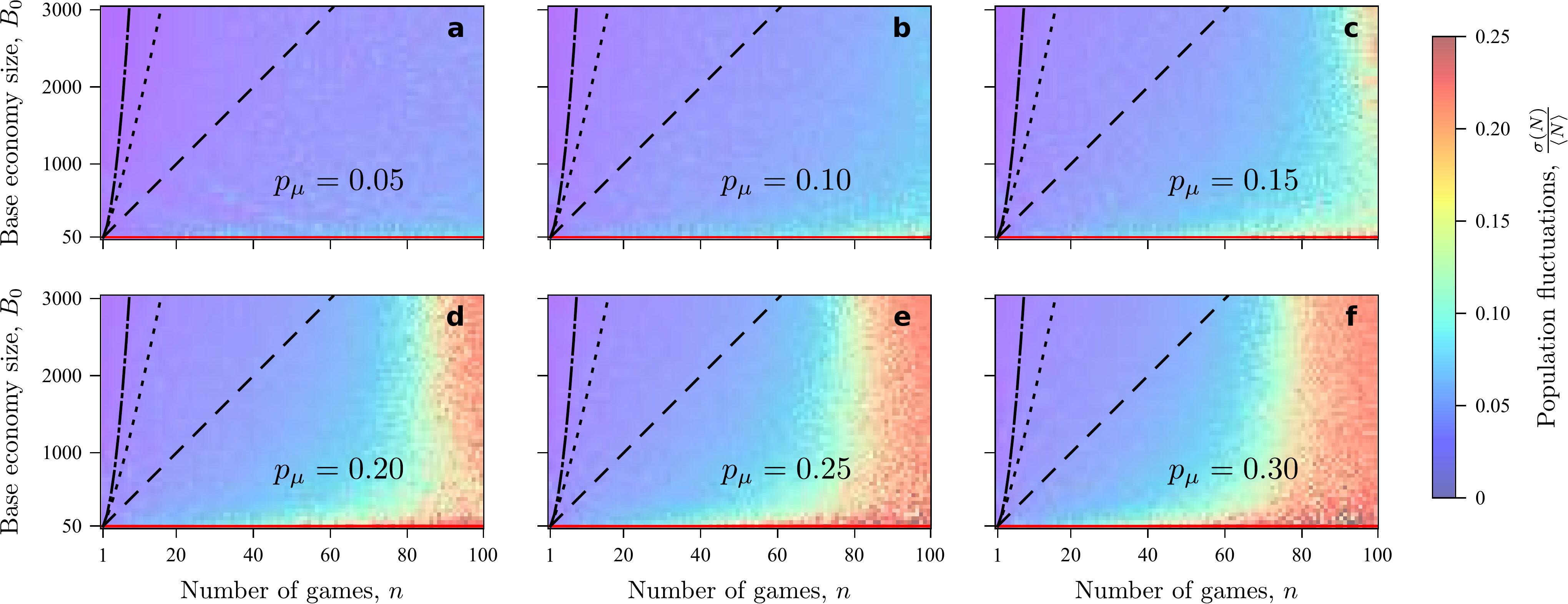}

        \caption{{\bf Comprehensive maps of $\sigma(N) / \left< N \right>$ for varying mutation, $p_\mu$. } Maps result
from simulating case $I$ (i.e. $B_I = B_0$, $b_I = B_0/n$) for ranges of economic complexity, $n$, and distributed
wealth, $B_0$. Black curves represent trajectories $B = B(B_0=50, n)$ for cases $II$, $III$, and $IV$ (dashed, dotted,
and dot-dashed respectively). Horizontal red lines at the bottom of each map represent case $I$ with $B_0 = 50$. {\bf
a}, $p_\mu = 0.05$; {\bf b}, $p_\mu = 0.10$; {\bf c}, $p_\mu = 0.15$; {\bf d}, $p_\mu = 0.20$; {\bf e}, $p_\mu = 0.25$;
{\bf f}, $p_\mu = 0.30$. }

        \label{fig:SI9}
      \end{center}
    \end{figure*}

    Supporting Figure \ref{fig:SI9} shows, indeed, that the mutation rate $p_\mu$ prompts this expected outcome too. As
    for $C_C$, $\left<N\right>$ is mostly unaffected by variations of $p_\mu$ (not shown), but increasing $p_\mu$ has
    the predicted result of advancing the onset of the large-fluctuations regime. In Supporting Figure \ref{fig:SI9} we
    show a range of $p_\mu$ with $C_C=20$. This choice of $C_C$ is different from the value taken in simulations in the
    main text ($C_C=10$) to show that, with a large enough $p_\mu$, we can advance the onset of large fluctuations to
    the point where it was with $C_C = 10$ and $p_\mu = 0.1$. 

    \begin{figure*}
      \begin{center}
        \includegraphics[width=\textwidth]{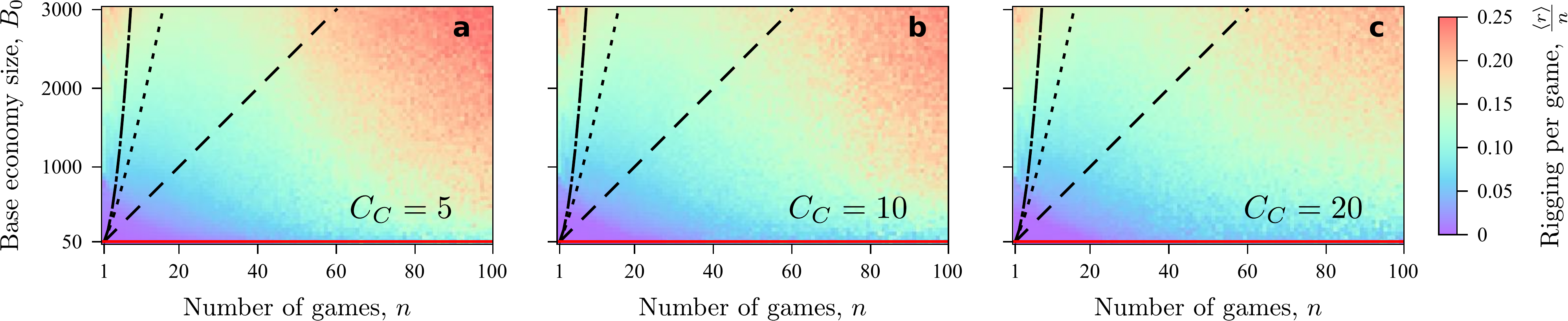}

        \caption{{\bf Comprehensive maps of rigging pressure per game for varying replication cost, $C_C$. } Maps result
from simulating case $I$ (i.e. $B_I = B_0$, $b_I = B_0/n$) for ranges of economic complexity, $n$, and distributed
wealth, $B_0$. Black curves represent trajectories $B = B(B_0=50, n)$ for cases $II$, $III$, and $IV$ (dashed, dotted,
and dot-dashed respectively). Horizontal red lines at the bottom of each map represent case $I$ with $B_0 = 50$. {\bf
a}, $C_C=5$; {\bf b}, $C_C=10$; {\bf c}, $C_C=20$. }

        \label{fig:SI10}
      \end{center}
    \end{figure*}

    Supporting Figure \ref{fig:SI10} shows the effect of $C_C$ on rigging pressure and rigging pressure per game. The
    effect is mostly uninteresting, as it just smooths or slightly displaces a map similar to the settings commented in
    the main text (Supporting Figure \ref{fig:SI8}). The key outcome is that rigging pressure grows both if more
    resources, $B$, and more games, $n$, are available -- as discussed in the main text. A similar structure is found
    for different values of $p_\mu$ (not shown). The strategy entropy (not shown) is largely unchanged by $C_C$ and
    $p_F$ -- while large enough $p_F$ (i.e. very large noise) has the expected effect of weakening the convergence to a
    majority strategy for low $n$ and large $B$ (arrows in Supporting Figure \ref{fig:SI8}{\bf a}). \\

    \begin{figure*}
      \begin{center}
        \includegraphics[width=\textwidth]{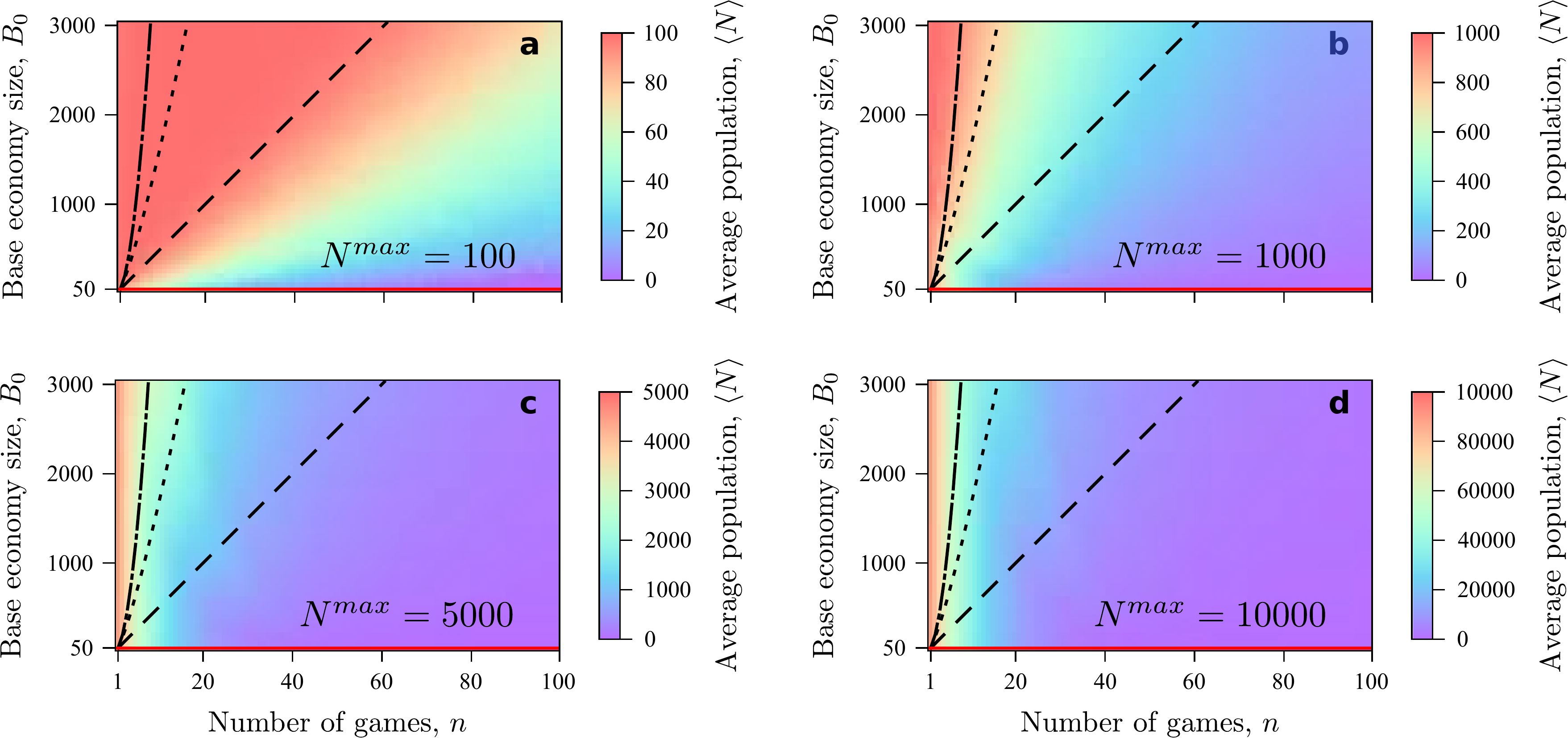}

        \caption{{\bf Comprehensive maps of $\left< N \right>$ for varying maximum population size, $N^{max}$. } Maps
result from simulating case $I$ (i.e. $B_I = B_0$, $b_I = B_0/n$) for ranges of economic complexity, $n$, and
distributed wealth, $B_0$. Black curves represent trajectories $B = B(B_0=50, n)$ for cases $II$, $III$, and $IV$
(dashed, dotted, and dot-dashed respectively). Horizontal red lines at the bottom of each map represent case $I$ with
$B_0 = 50$. {\bf a}, $N^{max} = 100$; {\bf b}, $N^{max} = 1000$; {\bf c}, $N^{max} = 5000$; {\bf d}, $N^{max} = 10000$.
}

        \label{fig:SI14}
      \end{center}
    \end{figure*}

    The model is a bit more sensitive to the parameter $N^{max}$ that sets an external maximum size to the population.
    This parameter makes sense as population size might be constrained by actual physical limits -- e.g., the amount of
    people that can occupy a territory. It could also be seen as a manufactured (`rigged') limit to the size of the
    market. This is an important kind of economic manipulation not studied in the model. Even if we were to look at
    $N^{max}$ from this perspective, model agents cannot modify it, so we would not be studying such manipulation
    organically, as we do with other degrees of freedom. 

    A more technical reason to set a parameter $N^{max}$ is that it solves parsimoniously the problem of a maximum
    average lifespan. In the limit $N^{max} \rightarrow \infty$, the first agent (who does not rig any game, hence does
    not pay $C_R$) never dies because it never ends up with negative wealth. The same would happen to any descent that
    does not rig any games. This is unrealistic and undesired. In the current model, agents that last too long are
    naturally and randomly replaced by newborns thanks to the finite $N^{max}$. If we would set an infinite $N^{max}$,
    we would need to introduce other mechanisms to remove unrealistically long-lasting agents -- e.g. an average
    life-time or removing agents with a wealth below a threshold $w^- > 0$. It would be interesting to try these and
    other variations in the future -- noting that they introduce new parameters nevertheless. 

    Ideally, we would like to find intrinsic limits to population size that emerge out of the model dynamics alone. To
    achieve this, we would need to simulate the model in the large $N^{max}$ limit.  The value chosen to report our
    results ($N^{max} = 1000$) is a compromise between a fairly large maximum population size and an ability to run
    simulations within a reasonable time. We ran one additional simulation for $N^{max} = 5000$ and another one for
    $N^{max} = 10000$ (reported next). Each of these took more than ten days in a fairly powerful computer cluster. 

    Supporting Figure \ref{fig:SI14} shows $\left<N\right>$ for $N^{max} = 100, 1000, 5000, 10000$. Some additional
    structure seems to emerge for intermediate values of $n$ (around $n \sim 20$) and $B$ (around $B \in [1000, 2000]$)
    for large $N^{max}$. It might be interesting to look at this, but it is not so relevant for our discussion. We see
    that the region associated to the large-fluctuations regime for large $n$, one of the most salient features of the
    model, is not very much affected by $N^{max}$. Average population sizes in the steady state in this regime (Figure
    \ref{fig:5}{\bf a}) as well as its fluctuations (Figure \ref{fig:5}{\bf b-c}) fall well below the chosen $N^{max} =
    1000$. All this suggests that the reported $\left<N\right>$ are  intrinsic limits emerging from the model. 

    \begin{figure*}
      \begin{center}
        \includegraphics[width=0.8\textwidth]{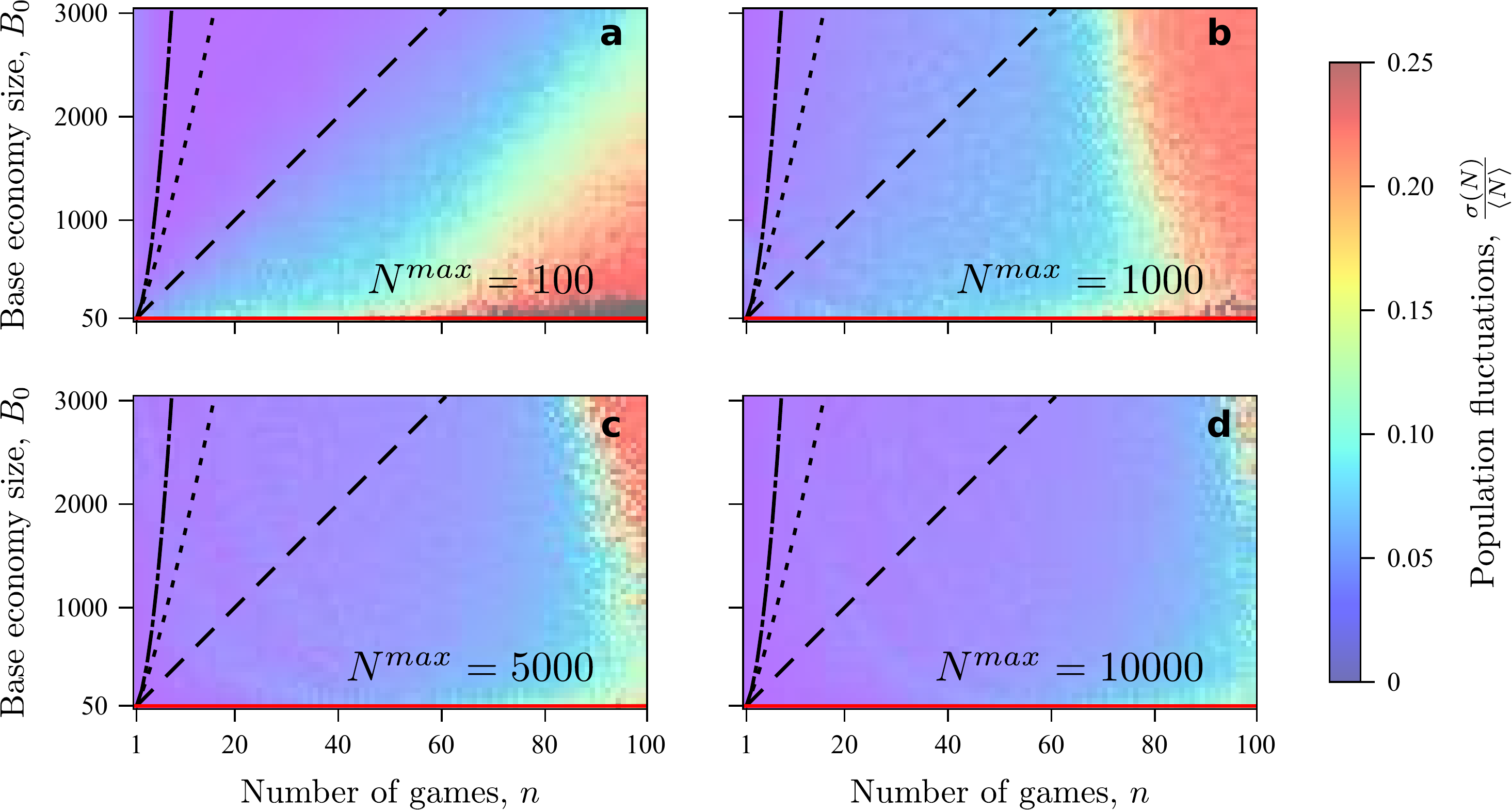}

        \caption{{\bf Comprehensive maps of $\sigma(N) / \left< N \right>$ for varying maximum population size,
$N^{max}$. } Maps result from simulating case $I$ (i.e. $B_I = B_0$, $b_I = B_0/n$) for ranges of economic complexity,
$n$, and distributed wealth, $B_0$. Black curves represent trajectories $B = B(B_0=50, n)$ for cases $II$, $III$, and
$IV$ (dashed, dotted, and dot-dashed respectively). Horizontal red lines at the bottom of each map represent case $I$
with $B_0 = 50$. {\bf a}, $N^{max} = 100$; {\bf b}, $N^{max} = 1000$; {\bf c}, $N^{max} = 5000$; {\bf d}, $N^{max} =
10000$. }

        \label{fig:SI15}
      \end{center}
    \end{figure*}

    The onset of the large-fluctuations regime, however, is more affected (Supporting Figure \ref{fig:SI15}). Increasing
    $N^{max}$ results in a displaced onset of this regime. It happens for notably larger values of $n$ and $B$ for
    increasing $N^{max}$. Within the plotted range, however, we still appreciate fluctuations as large as $25\%$ of
    population size for some $(n, B)$ combinations. Note that for $N^{max} = 10000$ the maximum population size is
    around two orders of magnitude bigger than the stable population size ($\sim 100$ for the affected area). This
    strongly suggests that large-fluctuations are an intrinsic regime of the model, even if its precise location in the
    map is affected by $N^{max}$. 

    Nevertheless, it is affected by $N^{max}$, which indicates that this phenomenon is enhanced by having a finite,
    maximum population size. Let us compare the shift of this onset with $N^{max}$ to the shifts observed when we varied
    $C_C$ (Supporting Figure \ref{fig:SI8.5}) and $p_\mu$ (Supporting Figure \ref{fig:SI9}). About these, we argued that
    the onset of the regime was advanced by mechanisms that result in more diverse strategies competing closer together
    or a higher population turnover. Thus, lower $C_C$ (cheaper reproduction) and higher $p_\mu$ (increased mutation)
    both advanced the onset of the regime because an array of diverse strategies is promptly forced to compete,
    potentially altering the winning rules in an unpredicted fashion. A larger $N^{max}$ has the effect of diluting this
    competence because there is less replacement of older agents. Accordingly, higher $N^{max}$ displaces the
    large-fluctuations regime to larger values of $n$ and $B$. Oppositely, note that the population renewal introduced
    by smaller $N^{max}$ results in a higher uncertainty about winning strategies. This is consistent, as discussed in
    the main text, with an onset of large fluctuations associated to a cognitive transition of the population as a
    whole: at some point, it becomes cognitively impossible to keep track of winning strategies as agents attempt to
    deceive each other. 

    \begin{figure*}
      \begin{center}
        \includegraphics[width=0.8\textwidth]{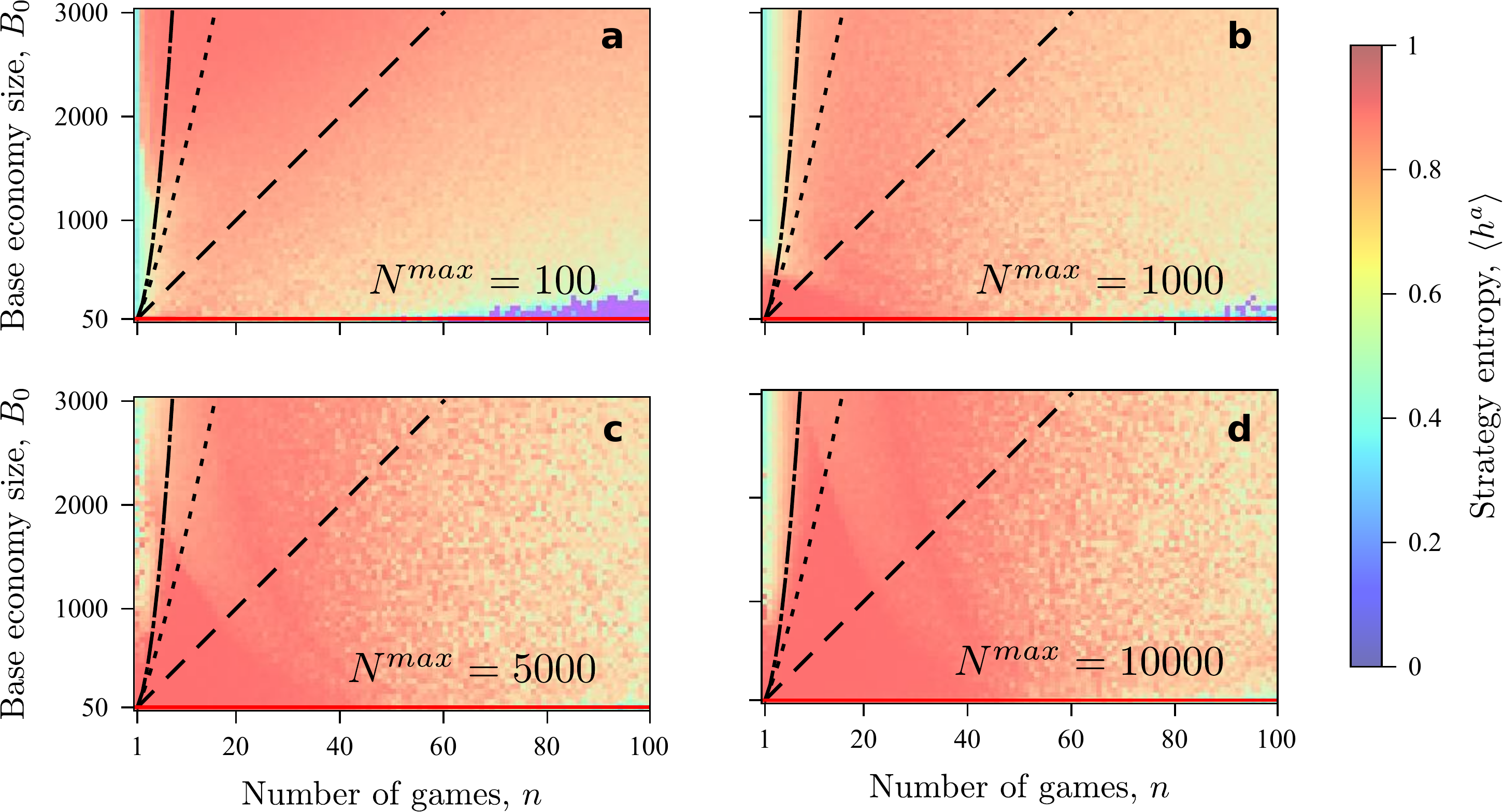}

        \caption{{\bf Comprehensive maps of strategy entropy, $\left< h^a \right>$ for varying maximum population size,
$N^{max}$. } Maps result from simulating case $I$ (i.e. $B_I = B_0$, $b_I = B_0/n$) for ranges of economic complexity,
$n$, and distributed wealth, $B_0$. Black curves represent trajectories $B = B(B_0=50, n)$ for cases $II$, $III$, and
$IV$ (dashed, dotted, and dot-dashed respectively). Horizontal red lines at the bottom of each map represent case $I$
with $B_0 = 50$. {\bf a}, $N^{max} = 100$; {\bf b}, $N^{max} = 1000$; {\bf c}, $N^{max} = 5000$; {\bf d}, $N^{max} =
10000$. }

        \label{fig:SI16}
      \end{center}
    \end{figure*}

    Another sensible area of the map is that with large $B$ and small $n$, which showed the transition to within-game
    coordination that results in most of the population converging to a same strategy for each game. Supporting Figure
    \ref{fig:SI14} shows that for such combination of parameters (large $B$ and small $n$), there are enough resources
    to sustain a very large average population size, often saturating even for large $N^{max}$ values. Supporting Figure
    \ref{fig:SI16} shows that this homogeneous regime remains present as we increase $N^{max}$. Some isolated cases in
    $N^{max}=10000$ (Supporting Figure \ref{fig:SI16}{\bf d}) and, more notably, $N^{max}=5000$ (Supporting Figure
    \ref{fig:SI16}{\bf c}) show up. These cases appear in the plots as outstanding pixels of huge $\left<h^a\right>$ in
    their otherwise smoother neighborhood with lower entropy (further supporting that these are oddballs). \\ 

    All in all, the results summarized in this appendix strongly suggest that the relevant phenomenology of the model is
    the one reported in the main text, and that this phenomenology is robust against reasonable variations of all
    parameters. Furthermore, changes on the onset of this phenomenology as we vary these extra parameters are
    parsimonious and follow logical explanations. All this, once again, is a strong reassurance of the minimalism of the
    model. This phenomenology likely underlies more complicated models that could study additional effects such as those
    discussed in the final section of the paper. We would expect to find at least similar regimes and regime shifts to
    the ones described here, even if the exact numerical values at which they happen are altered.


\end{document}